\documentclass[prd,singlecolumn,amsmath,nobibnotes,nofootinbib,superscriptaddress,preprintnumbers]{revtex4}

\usepackage{bm} \usepackage{graphicx}
\usepackage{color} \usepackage{amssymb}
\usepackage{epstopdf}
\usepackage{tikz}
\usepackage{rotating}
\usepackage{url}
\usepackage{multirow}

  \newcommand{\fsky}
  {f_{\rm{sky}}}

\newcommand{\Planck}
  {{\em{Planck}}}

\newcommand{\omk}
  {\Omega_k^{\rm obs}}
\newcommand{\ro}
  {r^{\rm obs}}
\newcommand{\rc}
  {r^{\rm coll}}
\newcommand{\ho}
  {H_{I}^{\rm obs}}
\newcommand{\hc}
  {H_{I}^{\rm coll}}
\newcommand{\afull}
  {2 \sqrt{\frac{8 \omk}{\ro}}}
\newcommand{\bfull}
  {2 \omk \sqrt{\frac{\rc}{\ro}}}
\newcommand{\bpfull}
  {2 \omk}

\newcommand{\alm}
  {a_{\ell m}}
\newcommand{\nlm}
  {n_{\ell m}}
\newcommand{\dlm}
  {d_{\ell m}}
\newcommand{\tlm}
  {t_{\ell m}}
\newcommand{\blm}
  {b_{\ell m}}
\newcommand{\tlz}
  {t_{\ell 0}}
\newcommand{\blz}
  {b_{\ell 0}}
\newcommand{\cell}
  {C_{\ell}}
\newcommand{\nell}
  {N_{\ell}}
\newcommand{\mean}
  {{\boldsymbol \mu}}
\newcommand{\cov}
  {{\mathbf C}}
\newcommand{\fisher}
  {{\mathbf F}}

\newcommand{\xc}
  {x_{\rm c}}
\newcommand{\thetac}
  {\theta_{\rm c}}
\newcommand{\xls}
  {x_{\rm ls}}
\newcommand{\dphi}
  {\frac{\delta\phi_0^{\rm coll}}{M_{\rm Pl}}}
\newcommand{\dxs}
  {\Delta x_{\rm sep}}
\newcommand{\cdxs}
  {1-\cos\dxs}
\newcommand{\rzero}
  {{\mathcal R}_0}
\newcommand{\rzerolin}
  {{\mathcal R}_0^L}
\newcommand{\rzeroquad}
  {{\mathcal R}_0^Q}
\newcommand{\linx}
  {\frac{x-\xc}{\xls}}
\newcommand{\quadx}
  {\frac{(x-\xc)^2}{\xls^2}}

\newcommand{\normlinlm}
  {\hat{L}_{\ell m}}
\newcommand{\normquadlm}
  {\hat{Q}_{\ell m}}
\newcommand{\xcnorm}
  {\frac{\xls}{\xls-\xc}}
\newcommand{\xcnorminv}
  {\frac{\xls-\xc}{\xls}}

\newcommand{\angs}
  {\omega}

\DeclareRobustCommand{\hatchleft}{%
\begin{tikzpicture}
\draw (0.8ex, 0.0)   -- (0.0, 0.8ex);
\draw (1.6ex, 0.0)   -- (0.0, 1.6ex);
\draw (1.6ex, 0.8ex) -- (0.8ex, 1.6ex);
\end{tikzpicture}
}

\DeclareRobustCommand{\hatchright}{%
\begin{tikzpicture}
\draw (0.8ex, 0.0)   -- (1.6ex, 0.8ex);
\draw (0.0,   0.0)   -- (1.6ex, 1.6ex);
\draw (0.0,   0.8ex) -- (0.8ex, 1.6ex);
\end{tikzpicture}
}

\begin{document}

\title{Forecasting constraints from the cosmic microwave background on eternal inflation}
\date{\today}

\author{Stephen M. Feeney}
\email{s.feeney@imperial.ac.uk}
\affiliation{Astrophysics Group, Imperial College London, Blackett Laboratory, Prince Consort Road, London, SW7 2AZ, U.K.}
\author{Franz Elsner}
\email{f.elsner@ucl.ac.uk }
\affiliation{Department of Physics and Astronomy, University College London, London WC1E 6BT, U.K.}
\author{Matthew C. Johnson}
\email{mjohnson@perimeterinstitute.ca}
\affiliation{Perimeter Institute for Theoretical Physics, Waterloo, Ontario N2L 2Y5, Canada} 
\affiliation{Department of Physics and Astronomy, York University, Toronto, Ontario, M3J 1P3, Canada}
\author{Hiranya V. Peiris}
\email{h.peiris@ucl.ac.uk}
\affiliation{Department of Physics and Astronomy, University College London, London WC1E 6BT, U.K.}

\begin{abstract}

We forecast the ability of cosmic microwave background (CMB) temperature and polarization datasets to constrain theories of eternal inflation using cosmic bubble collisions. Using the Fisher matrix formalism, we determine both the overall detectability of bubble collisions and the constraints achievable on the fundamental parameters describing the underlying theory. The CMB signatures considered are based on state-of-the-art numerical relativistic simulations of the bubble collision spacetime, evolved using the full temperature and polarization transfer functions. Comparing a theoretical cosmic-variance-limited experiment to the WMAP and \Planck\ satellites, we find that there is no improvement to be gained from future temperature data, that adding polarization improves detectability by approximately 30\%, and that cosmic-variance-limited polarization data offer only marginal improvements over \Planck. The fundamental parameter constraints achievable depend on the precise values of the tensor-to-scalar ratio and energy density in (negative) spatial curvature. For a tensor-to-scalar ratio of $0.1$ and spatial curvature at the level of $10^{-4}$, using cosmic-variance-limited data it is possible to measure the width of the potential barrier separating the inflating false vacuum from the true vacuum down to $M_{\rm Pl}/500$, and the initial proper distance between colliding bubbles to a factor $\pi/2$ of the false vacuum horizon size (at three sigma). We conclude that very near-future data will have the final word on bubble collisions in the CMB.
\end{abstract}

\preprint{}

\maketitle


\section{Introduction}

Imminent results from the \Planck\ satellite~\cite{Tauber2010} will contain nearly all of the large-scale cosmological information encoded in the temperature and polarization anisotropies of the cosmic microwave background (CMB) radiation. This makes the present a highly opportune time to determine which analyses can maximize the scientific return from this dataset. One exciting opportunity is the potential observation of relics left from events that occurred in the very early Universe. In this paper, we forecast the ability of a cosmic-variance-limited CMB dataset to constrain one such class of early Universe events: cosmic bubble collisions in eternal inflation.

Inflation, a hypothesized epoch of accelerated expansion, has become an essential component of the standard cosmological model \citep{1982PhLB..115..295H,1982PhLB..117..175S,1982PhRvL..49.1110G,1983PhRvD..28..679B}. As a side-effect, inflation can in many cases give rise to an eternally inflating multiverse \citep{steinhardt1982,1983PhRvD..27.2848V}. In this scenario, a high-energy inflating phase is exited locally, inside bubbles, but not globally. In eternal inflation, the rate of bubble formation is outpaced by the accelerated expansion of the high-energy inflating phase, preventing the percolation of bubbles and leaving an increasingly large volume in which inflation and bubble formation continues. An important test of this scenario is the observation of the wreckage left from collisions between our own bubble and others~\cite{Aguirre:2007an}. Determining the outcome of collisions and the probability of observing them has been the subject of a substantial body of work~\cite{Hawking:1981fz,Hawking:1982ga,Wu:1984eda,Moss:1994pi,Aguirre:2007wm,Aguirre:2008wy,Aguirre:2009ug,Kozaczuk:2012sx,Chang_Kleban_Levi:2009,Chang:2007eq,Czech:2010rg,Freivogel_etal:2009it,Gobbetti_Kleban:2012,Kleban_Levi_Sigurdson:2011,Kleban:2011pg,Wainwright:2013lea,Johnson:2010bn,Johnson:2011wt,Freivogel:2007fx,Wainwright:2014pta,Salem:2012gm,Czech:2011aa,Salem:2011qz,Salem:2010mi,Larjo:2009mt,Easther:2009ft,Giblin:2010bd,Ahlqvist:2014uha,Kim:2014ara,Ahlqvist:2013whn,Hwang:2012pj,Deskins:2012tj,Amin:2013dqa,Amin:2013eqa}. Observational constraints on this scenario have already been placed using temperature data from the {\it Wilkinson Microwave Anisotropy Probe} (WMAP) satellite~\cite{Feeney_etal:2010dd,Feeney_etal:2010jj,Feeney:2012hj,McEwen:2012uk,Osborne:2013hea,Osborne:2013jea}. 

\begin{figure}[h!]
		\begin{center}
			\includegraphics[width=8cm]{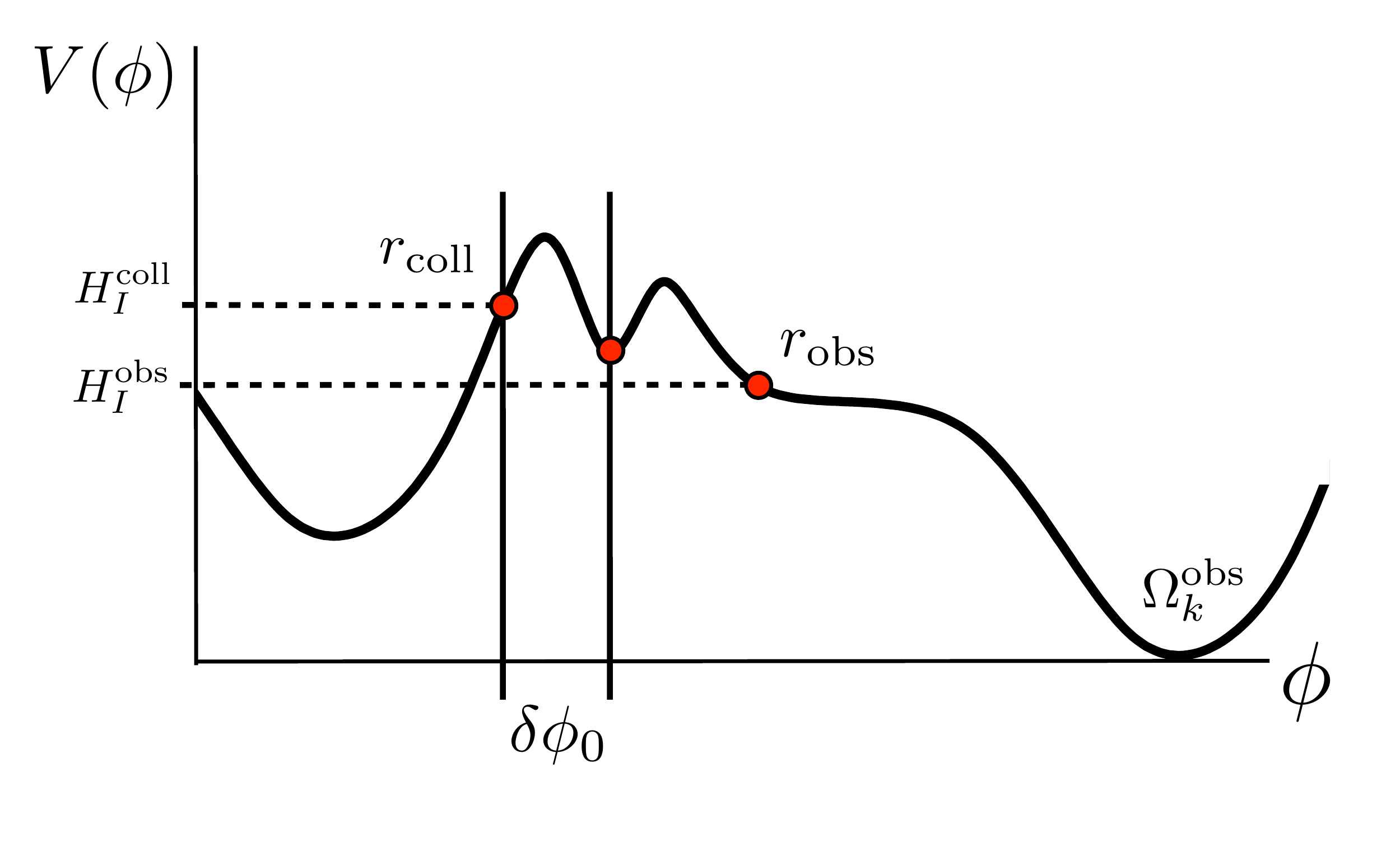}
			\caption{An example of a potential giving rise to eternal inflation with two possible types of bubbles. The components of the potential determining the various parameters in Eq.~\ref{eq:raw_template} are labeled.}
			\label{fig:potential}
		\end{center}
\end{figure}

There have been a number of important developments since the early work on constraining bubble collisions with observations. Importantly, a direct link has been made between the scalar field Lagrangian and cosmological observables~\cite{Wainwright:2013lea,Wainwright:2014pta}. For models containing a single scalar field with a canonical kinetic term, the properties of the eternally inflating Universe are determined entirely by the potential of the scalar field. An example is shown in  
 Fig.~\ref{fig:potential}; this model allows two types of bubbles. Because a bubble collision gives rise to an inhomogeneous Universe, observers at different locations will have access to different parts of the collision spacetime. For observers near the causal boundary of the collision there is a simple analytic template for the comoving curvature perturbation caused by a single bubble collision in single-scalar-field models~\cite{Wainwright:2014pta}, given by
 \begin{equation}\label{eq:raw_template}
\mathcal{R} = \afull \dphi \left( \cdxs \right)  \linx + \bfull \frac{\hc}{\ho} \left( \cdxs\right)^2 \quadx \, ,
\end{equation}
where $\omk$ is the curvature inside the bubble containing the observers, $\ro$ and $\rc$ are the inflationary tensor-to-scalar ratios inside the observation and collision bubbles, $\ho$ and $\hc$ are the Hubble scales during inflation inside the observation and collision bubbles, $\delta\phi_0^{\rm coll}$ is the width of the potential barrier separating the inflating false vacuum from the true vacuum in the collision bubble, $M_{\rm pl}$ is the Planck mass, $0 < \dxs < \pi$ is the distance between the colliding bubbles (measured in terms of the false-vacuum Hubble parameter) in the centre of mass frame, $\xc$ is the comoving position of the causal boundary of the collision, and $\xls$ is the comoving distance to the surface of last scattering. The parameter $\dxs$ varies from collision to collision, and $\xc$ depends on the position of the observer; all other parameters are fixed by the scalar field Lagrangian as shown in Fig.~\ref{fig:potential}. Eq.~\ref{eq:raw_template} is valid in the limit where the slow-roll
approximation holds in the future of the collision inside both the observation and collision bubbles. Outside this limit, numerical simulations are necessary to accurately determine the template. All cases considered below are well described by Eq.~\ref{eq:raw_template}.

\begin{figure}
		\begin{center}
			\includegraphics[width=7.5cm]{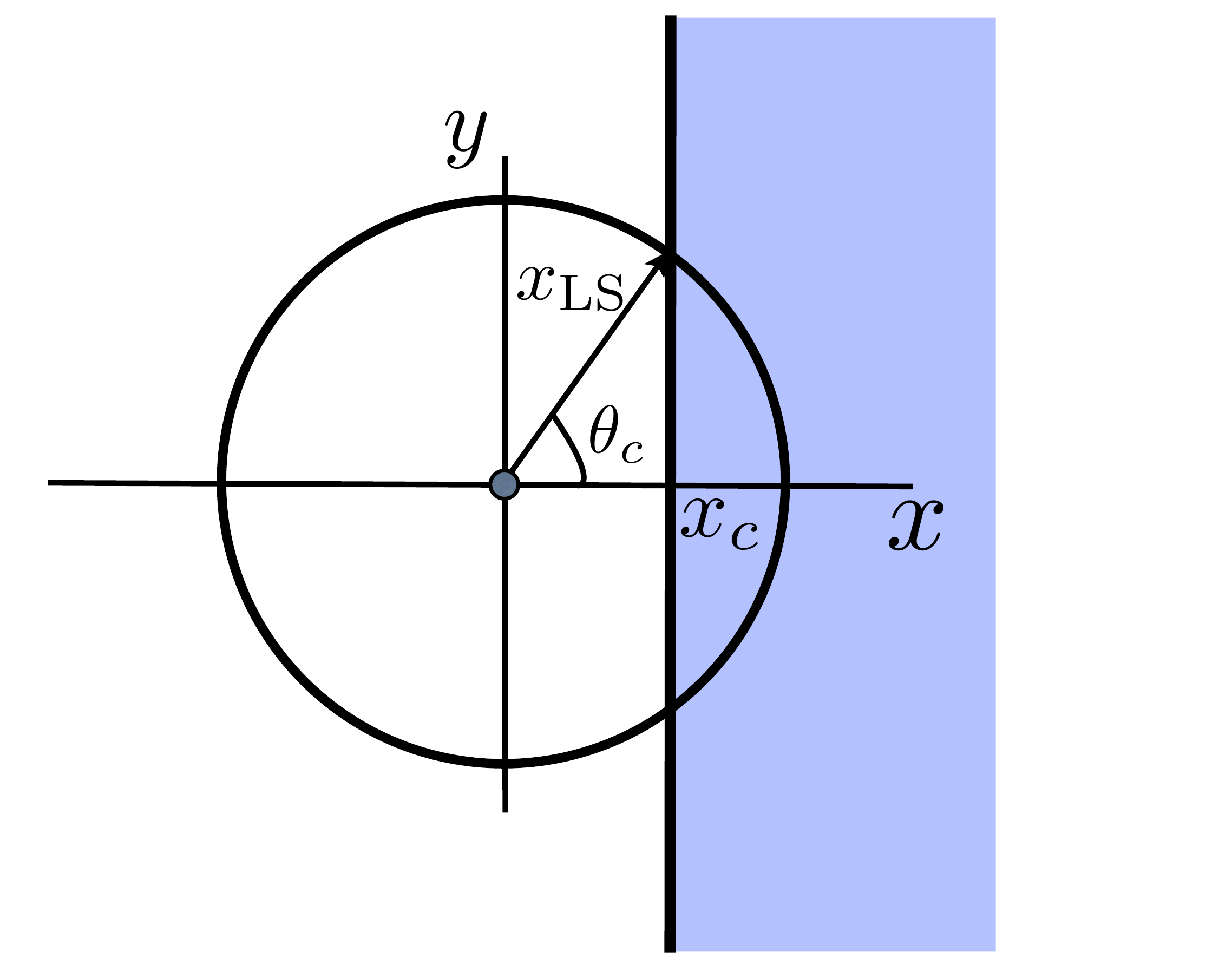}
			\includegraphics[width=7.5cm]{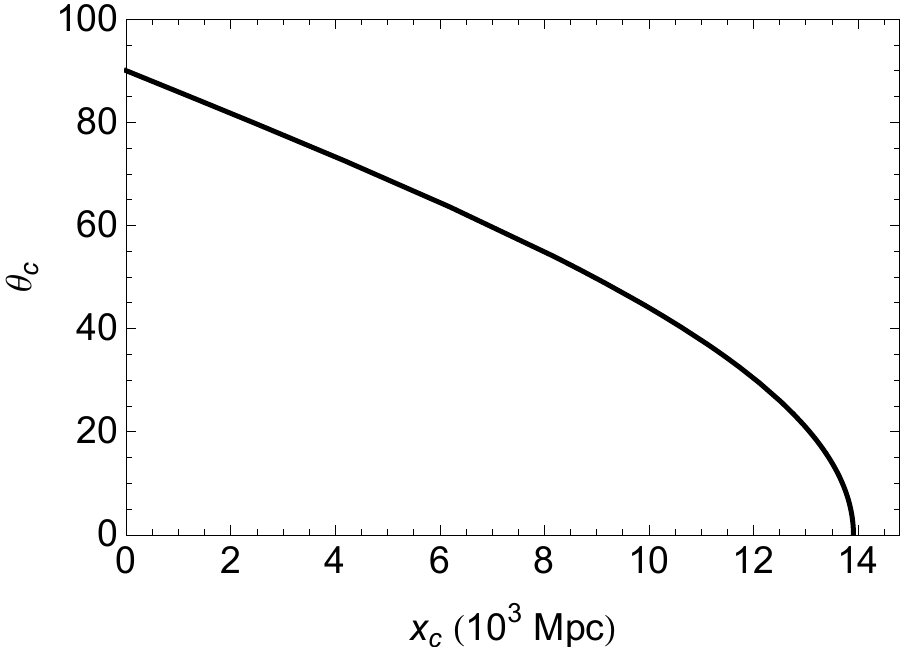}
			\caption{In the left panel, we show the geometry of a bubble collision on the surface of last scattering. Shaded regions are affected by the collision, while unshaded regions are not. The observer is located at the origin, and the circle represents their past light cone. In the right panel, we show the mapping between the observed angular radius of the collision $\thetac$ in degrees and the comoving position of the collision boundary $\xc$ in Mpc.}
			\label{fig:xc_to_angle}
		\end{center}
\end{figure}

The collision spacetime possesses SO(2,1) symmetry, which translates into an approximate planar symmetry in the neighbourhood of an observer in the limit $ \Omega_k^{\rm obs} \ll 1$. This geometry is shown in Fig.~\ref{fig:xc_to_angle}. The planar symmetry of the collision and the existence of a causal boundary separating the bubble into regions that are or are not affected by the collision imply that the effects of the collision are confined to a disc of angular radius $\thetac$ on the sky of an observer. The mapping between the position of the causal boundary $\xc$ (in Mpc) and the angular scale of the collision for a frame in which the observer is at the origin of coordinates is shown in the right panel of Fig.~\ref{fig:xc_to_angle}.\footnote{In generating this plot, and throughout the paper, we have assumed the 2013 \Planck+WP+highL+BAO best fit cosmology~\cite{2014A&A...571A..16P}.}

In this paper, we forecast the ability of a suite of CMB experiments, producing both temperature and polarization data, to detect bubble collisions of the form given in Eq.~\ref{eq:raw_template}. We forecast constraints on the overall detectability of bubble collisions as well as on the fundamental parameters underlying the collision model. Although constraints~\cite{Feeney_etal:2010dd,Feeney_etal:2010jj,Feeney:2012hj,McEwen:2012uk,Osborne:2013hea,Osborne:2013jea} and forecasts~\cite{Kleban_Levi_Sigurdson:2011} exist for bubble collisions in the CMB, so far they have assumed a phenomenological and incomplete template. This paper is an important step forward as we forecast constraints directly on the scalar-field potential that underlies eternal inflation. As the relevant datasets become available, searches along the lines of Refs.~\cite{Feeney_etal:2010dd,Feeney_etal:2010jj,Feeney:2012hj,McEwen:2012uk,Osborne:2013hea,Osborne:2013jea} can be implemented using the complete collision template.

The paper is structured as follows. In Sec.~\ref{sec:evolving} we describe how the curvature perturbation template is evolved to the CMB observables. Sec.~\ref{sec:fisher} describes our forecasting technique, which is based on the Fisher information matrix. In Secs.~\ref{sec:results} and~\ref{sec:comparison} we present our forecasts and compare them to existing constraints; we conclude in Sec.~\ref{sec:discussion}.

\section{Evolving the primordial comoving curvature perturbation}\label{sec:evolving}

Given the analytical parameterization of the curvature perturbations in
Eq.~\ref{eq:raw_template}, we now describe the formalism adopted to
predict bubble signatures imprinted onto the CMB. The spherical
harmonic coefficients $b_{\ell m}$ of the template are related to the
curvature perturbations $\mathcal{R}_{\ell m}(k)$ via (e.g., Ref.~\cite{2003ApJS..148..119K})
\begin{equation}
  b^{X}_{\ell m}=\frac{(-\imath)^{\ell}}{2\pi^2} \int dk \, k^2 \,
   \Delta^{X}_{\ell}(k) \ \mathcal{R}_{\ell m}(k) \, ,
\end{equation}
where $\Delta^{X}_{\ell}(k)$ is the radiation transfer function in
temperature and polarization ($X = \{T, E\}$) in momentum space.\footnote{In our sign convention, the CMB temperature anisotropies in the Sachs-Wolfe approximation are related to the comoving curvature perturbation by $\delta T/T  = \mathcal{R} / 5$.} For an
efficient solution of the problem, it is more convenient to phrase the
equation in terms of comoving distances,
\begin{equation}
  \label{eq:r2alm}
  b^{X}_{\ell m} = \int dr \, r^2 \, \alpha_\ell^{X}(r)
  \ \mathcal{R}_{\ell m}(r) \, ,
\end{equation}
where we have introduced the real-space transform of the radiation transfer
function,
\begin{equation}
  \alpha_\ell^{X}(r) = \frac{2}{\pi} \int dk \, k^2 \, \Delta_\ell^{X}(k)
  \ j_\ell(kr) \, ,
\end{equation}
where the $j_\ell(kr)$ are spherical Bessel functions of order $\ell$.

To evaluate Eq.~\ref{eq:r2alm} numerically, we first obtain the
radiation transfer function using a modified version of the Boltzmann
integrator CAMB~\cite{2002PhRvD..66j3511L}. In doing so, we assume the 2013 
\Planck+WP+highL+BAO best-fit cosmological parameters~\cite{2014A&A...571A..16P}, adopting the
limiting case of a flat cosmology. Given our
parametrization of $\mathcal{R}_{\ell m}(r)$ as a function of $\linx$,
we then compute the radial integral on 90 nodes for all multipole
moments $\ell, |m| \le 2500$. We use the approach introduced in
Ref.~\cite{2009ApJS..184..264E} to optimize node positions and quadrature
weights to minimize the error associated with the numerical
integration. Resolution studies have proven results to be stable.

\begin{figure}
  \begin{center}
    \includegraphics[width=0.49\textwidth]{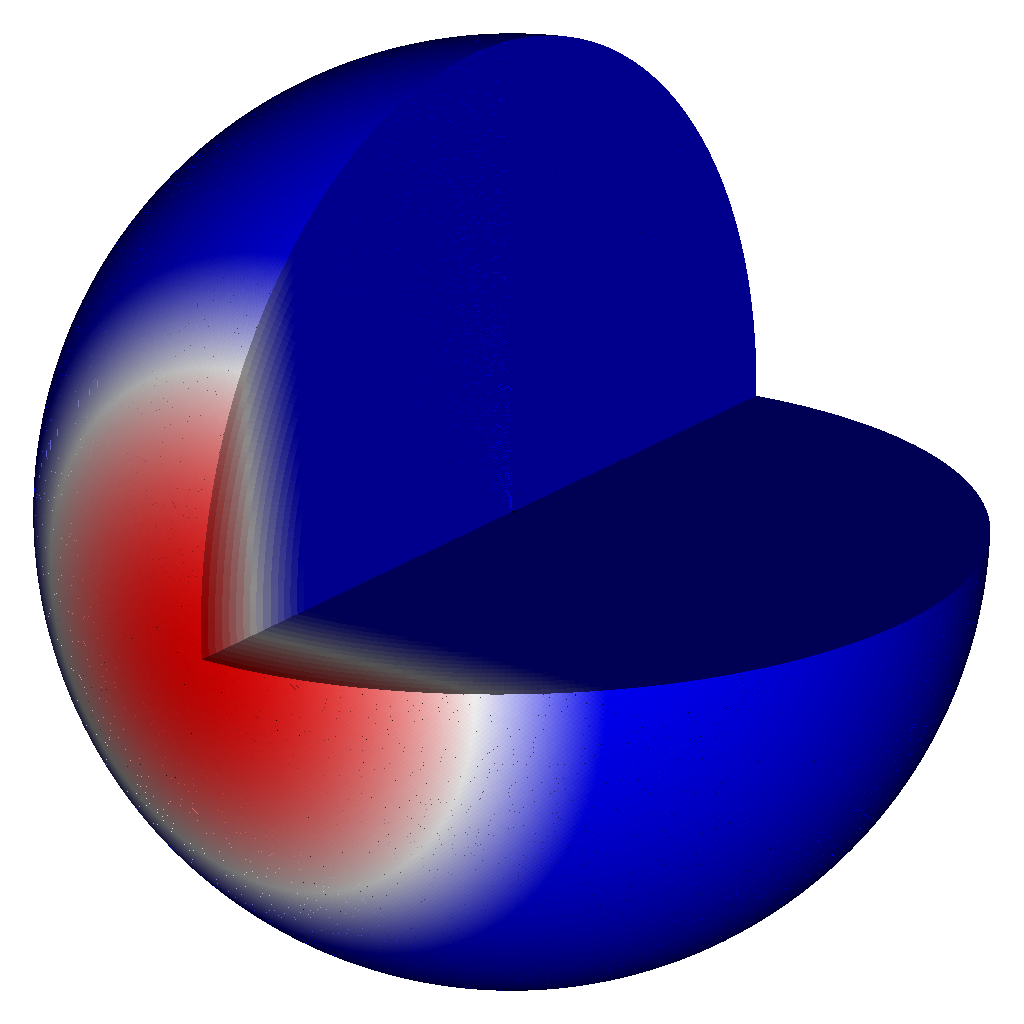}
    \includegraphics[width=0.49\textwidth]{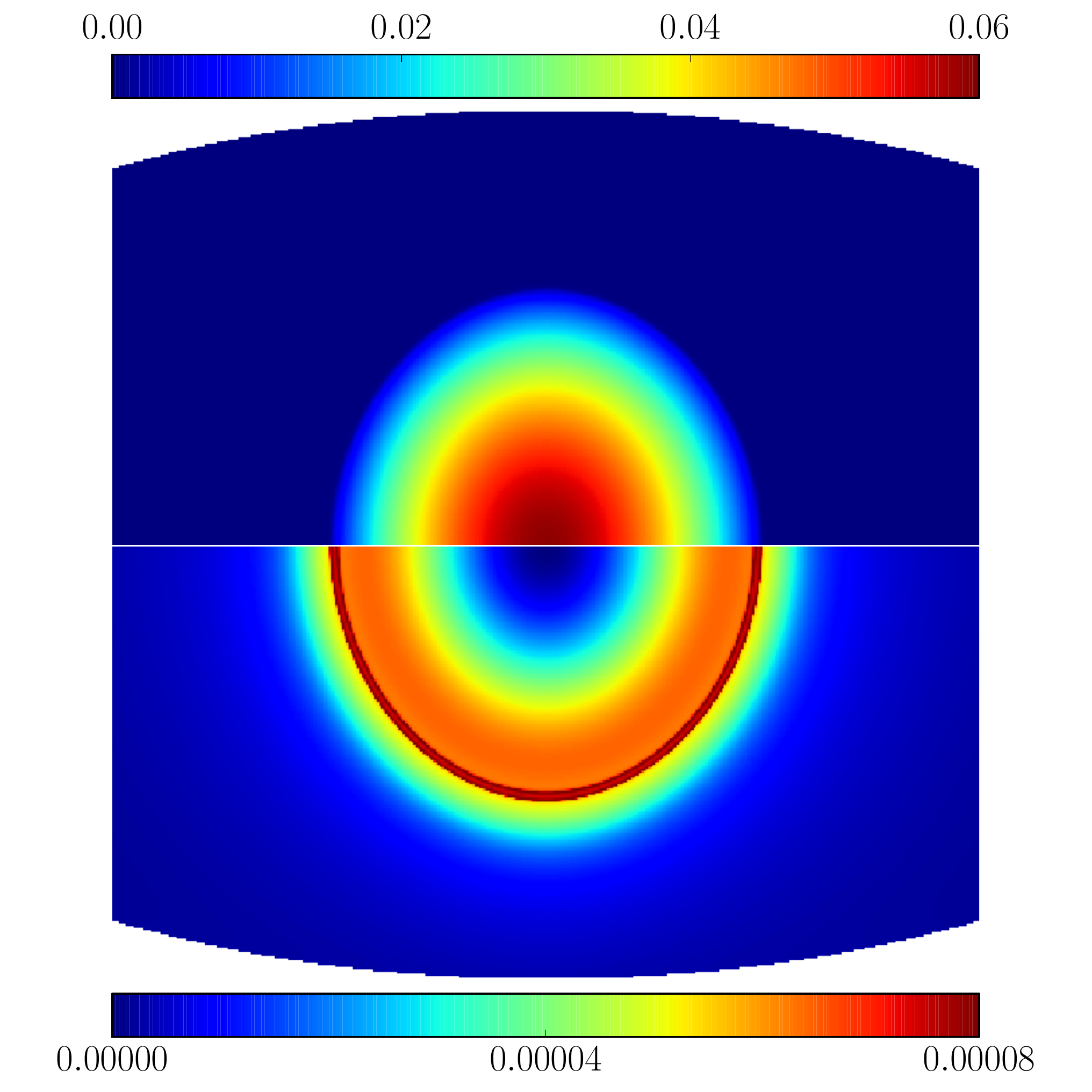}
    \caption{The induced signal in the three-dimensional curvature
      perturbation (\emph{left panel}, shown on shells centered on the
      observer out to the last scattering surface) results in radially
      symmetric features in the CMB. \emph{Right panel:} with the
      direction of the bubble collision in the center of the plot, we
      show Mollweide projections of the corresponding templates in
      temperature (\emph{upper half}) and polarization (\emph{lower
        half}) in dimensionless units.}
    \label{fig:r_example_visualization}
  \end{center}
\end{figure}

In Fig.~\ref{fig:r_example_visualization}, we visualize the outlined
procedure for the linear term $\mathcal{R} \propto x - \xc$, with the
distance to the causal boundary of the collision $\xc$ set to 10 Gpc. We
plot the curvature potential on a three-dimensional spherical grid
enclosing the observable Universe in comoving coordinates (left), and
juxtapose the derived templates for the temperature and polarization
intensity (defined by $P = \sqrt{Q^2 + U^2}$) signals (right). Whereas the
former is a relatively smooth function with a well defined boundary,
the latter shows a more extended signature and a comparatively sharp
feature at the location of the observed angular radius of the
collision. This feature is a result of the discontinuous first
derivative of the curvature perturbation at $x = \xc$~\cite{Kleban_Levi_Sigurdson:2011}, and is consequently absent in the
quadratic term, where $\mathcal{R} \propto (x - \xc)^2$.

In practice, we need only transform the two position-dependent terms $(x - \xc)$ and $(x - \xc)^2$ individually to build any template desired (at a given scale). We normalize the curvature perturbations so that they are fixed to 0 at $x=\xc$ and 1 at $x=\xls$,
and denote their evolved forms $\normlinlm^X(\xc, \angs)$ and $\normquadlm^X(\xc, \angs)$, respectively, where $\angs = \left( \theta, \phi \right)$ are the angular coordinates of the centre of the signature on the sphere. To minimize the number of parameter degeneracies we face, we restrict our analysis to collisions between identical bubbles, setting $\rc = \ro$ and $\hc = \ho$. In this case, the spherical harmonic coefficients of a bubble collision in the CMB temperature or polarization can be written as
\begin{align}
\blm^X = & A \dphi \left[ \cdxs \right] \left[ \xcnorm \right] \normlinlm^X(\xc, \angs) \, + \\
& B \left[ \cdxs \right]^2 \left[ \xcnorm \right]^2 \normquadlm^X(\xc, \angs),
\label{eq:template_lag}
\end{align}
where $X = \{T,E\}$ and the constants
\begin{equation}
A = \afull; \, B = \bpfull
\end{equation}
are assumed to have been constrained by other data. When presenting constraints on fundamental parameters, we assume the hypothetical situation where primordial gravitational waves and negative spatial curvature have been detected at a level of $\ro = 0.1$ and $\omk = 10^{-4}$, respectively, yielding the numerical values $A \simeq 0.179$ and $B = 2 \times 10^{-4}$. Throughout our analysis, and without loss of generality, we choose to centre the bubble collisions on the North Pole. This, combined with the azimuthal symmetry of the bubble collision signature, means that all $\blm^X$ with $m \ne 0$ are zero.


\section{Fisher matrix analysis}\label{sec:fisher}

Having obtained the signature of bubble collisions in the CMB, we now describe the method whereby we forecast constraints on the underlying parameters. We assume that the data --- in this case, the observed spherical harmonic coefficients $\dlm^X$, where $X = \{T,E\}$ --- consist of the stochastic Gaussian CMB ($\alm^X$), the beam-deconvolved instrumental white noise ($\nlm^X$), and a deterministic bubble collision ($\blm^X$). Under these assumptions the moments of the (Gaussian) data are simple to define: the noise and CMB anisotropies do not contribute to the mean, which is determined entirely by the collision to be
\begin{equation}
\langle {\mathbf d} \rangle = \mean = \begin{pmatrix} \blm^T \\ \blm^E \end{pmatrix};
\end{equation}
and the deterministic bubble collision signature does not contribute to the covariance, which is defined purely by the CMB and noise power spectra to be
\begin{equation}
\langle {\mathbf d} {\mathbf d}^t \rangle - \langle {\mathbf d} \rangle \langle {\mathbf d} \rangle^t = \cov = \begin{pmatrix} \cell^{TT} + \nell^{TT} & \cell^{TE} \\ \cell^{TE} & \cell^{EE} + \nell^{EE} \end{pmatrix}.
\label{eq:covariance}
\end{equation}
We generate the CMB power spectra $\cell$ using CAMB, fixing the cosmological parameters to their 2013 \Planck+WP+highL+BAO best-fit values~\cite{2014A&A...571A..16P}

We produce forecasts for three experimental configurations: a theoretical cosmic-variance-limited experiment with $\nell = 0$, a forecast of the \Planck\ satellite's full mission, and WMAP, included to compare our forecasts against existing results. Following Ref.~\citep{2014PhRvD..90f3504G}, we define the \Planck\ forecast to correspond to 30 months of observations, retaining only two CMB channels (143 and 217 GHz) to account for foreground subtraction. The beam-deconvolved noise power spectrum for this mission is
\begin{equation}
\nell^{XX} = \left( \sum_\nu \left[ \left( \sigma_\nu^X \theta_\nu \right)^2 \exp{ \left( \frac{\ell (\ell + 1) \theta_\nu^2 }{ 8 \ln 2 } \right) } \right]^{-1} \right)^{-1},
\end{equation}
where $\theta_\nu$ is the full-width at half-maximum of each channel's (Gaussian) beam in radians and $\left( \sigma_\nu^X \right)^2$ is the noise variance per beam-sized patch in temperature or polarization. Our WMAP experiment corresponds to the seven-year, foreground-reduced W-band {\em temperature} map~\cite{WMAP7_Jarosik}. The precise values of the experimental parameters are presented in Table~\ref{tab:experiments}. Note that we do not include the effects of partial sky coverage in our analysis, and the forecasts we present therefore apply to signatures that are not significantly masked.\footnote{The mode-count reduction due to masking can be approximated by inserting a factor of $\fsky^{-1}$ in the covariance matrix (Eq.~\ref{eq:covariance}). The net effect is to boost all parameter uncertainties by a factor of $\fsky^{-1/2}$. Note that the relevant sky fraction depends on the size of the collision and its position with respect to the galactic plane and other masked sources.}

\begin{table*}
\begin{tabular}{c c c c c}
\hline
\hline
Experiment & \ $\nu $ / GHz \ & \ $\sigma_\nu^T$ / $\mu$K \ & \ $\sigma_\nu^P$ / $\mu$K \ & \ $\theta_\nu$ / arcmin \\
\hline
\multirow{2}{*}{\Planck} & 143 & 4.1 & 7.8 & 7.1 \\
    & 217 & 8.9 & 18.2 & 5.0\\
WMAP & 94 & 35.3 & - & 13.2 \\
\hline
\hline
 \end{tabular}
 \begin{center}
 \caption{The experimental characteristics assumed in this work. 
   \label{tab:experiments}}
 \end{center}
\end{table*}

The scope of this work is restricted to investigating the parameters describing the bubble collisions ($\theta_i$), so the covariance is independent of the parameters of interest. In this setting, the Fisher matrix is given by~\cite{1997ApJ...480...22T}
\begin{equation}
\fisher_{ij} = - \left\langle \frac{ \partial^2 \ln {\mathcal L} }{ \partial \theta_i \partial \theta_j } \right \rangle = \frac{1}{2} \sum_{\ell m} {\rm Tr} \left( \cov^{-1} \left[ \frac{\partial\mean}{\partial\theta_i} \frac{\partial\mean^t}{\partial\theta_j} + \frac{\partial\mean}{\partial\theta_j} \frac{\partial\mean^t}{\partial\theta_i} \right] \right).
\end{equation}
Assuming the likelihood ${\mathcal L}$ is Gaussian in the bubble parameters, we can associate the inverse of the Fisher matrix with the covariance of the parameters
\begin{equation}
\fisher^{-1}_{ij} = \langle \theta_i \theta_j \rangle - \langle \theta_i \rangle \langle \theta_j \rangle.
\end{equation}
The diagonal of the inverse Fisher matrix corresponds to the variance on each parameter after having marginalized over all others; the inverse of each diagonal element of the Fisher matrix is the variance on each parameter conditional on all other parameters taking their maximum-likelihood values.

\subsection{Parameterizations and Detectability}

Our goals in this work are to forecast the detectability of bubble collision signatures as well as our ability to constrain the fundamental parameters describing them. Both of these goals can be achieved using the Fisher matrix formalism described above, simply by re-parameterizing the collision signature. The simplest starting point is to cast the analysis in terms of the Lagrangian parameters ($\boldsymbol{\theta} = \{\dphi, \dxs, \xc, \angs\}$). The resulting Fisher matrix encodes our ability to measure the fundamental parameters describing the bubble collision space-time, but does not yield a simple concept of detectability. To define our detectability criteria, we re-cast the templates in terms of observable amplitudes, and require that these amplitudes can be distinguished from zero with some threshold significance.

With this concept in mind, we employ two additional parameterizations in this work, based on observable rather than fundamental quantities. Specifically, we re-cast the template in terms of the individual amplitudes of the linear and quadratic terms ($\boldsymbol{\theta}^\prime = \{\rzerolin, \rzeroquad, \xc, \angs\}$), or the total amplitude and the fraction of the template contributed by the linear term ($\boldsymbol{\theta}^{\prime\prime} = \{\rzero, f, \xc, \angs\}$). The various amplitudes are defined by
\begin{align}
\blm^X & = \rzerolin \normlinlm^X(\xc, \angs) + \rzeroquad \normquadlm^X(\xc, \angs) \label{eq:template_obs_lq}\\
& = \rzero \left[ f \normlinlm^X(\xc, \angs) + (1-f) \normquadlm^X(\xc, \angs) \right],\label{eq:template_obs_tot}
\end{align}
and their parameters are related via
\begin{align}
\rzerolin & = A \dphi \left[ \cdxs \right] \left[ \xcnorminv \right] \label{eq:rzerolin_trans} \\
\rzeroquad & = B \left[ \cdxs \right]^2 \left[ \xcnorminv \right]^2 \label{eq:rzeroquad_trans}\\
\rzero & =  \rzerolin+ \rzeroquad. \label{eq:rzero_trans}
\end{align}
By determining at what point the overall ($\rzero$), linear ($\rzerolin$), and quadratic ($\rzeroquad$) amplitudes become distinguishable from zero at a given significance, we are able to state when mixed, pure-linear, and pure-quadratic bubble collision signatures can be detected.

To obtain constraints on the full range of parameters we need only calculate the Fisher matrix in terms of one set and transform the resulting inverse matrix using
\begin{equation}
\fisher(\boldsymbol{\theta}^\prime)^{-1}_{ij} = \sum_{kl} \frac{\partial \theta^\prime_i}{\partial \theta_k} \fisher(\boldsymbol{\theta})^{-1}_{kl} \frac{\partial \theta^\prime_j}{\partial \theta_l}.
\label{eq:fisher_transform}
\end{equation}
Fundamental to the Fisher matrix formalism is the assumption that the likelihood is Gaussian in the parameters of interest. This guarantees that the initial parameterization choice has no bearing on the Fisher matrices produced for each other parameterization; it also means that the resulting forecasts are more accurate for parameterizations which better satisfy this assumption. As the likelihood is bivariate Gaussian in the linear and quadratic amplitudes, we expect these to be most accurately predicted, whereas the error for the total amplitude, for example, will be larger.

\subsection{Implementation}

As our aim is to forecast our ability to detect and constrain the bubble collision model as a function of the Lagrangian parameters, we frame the Fisher matrix analysis in terms of this parameterization, with one small change: in order to linearize the dependence of the template on the parameters (and hence make the likelihood as close as possible to Gaussian in the parameters), we use $\cdxs$ instead of $\dxs$. From Eq.~\ref{eq:template_lag}, we see that the derivatives with respect to $\dphi$ and $\cdxs$ can be computed analytically. We calculate the derivatives with respect to $\xc$ numerically, using two-sided finite differences with a step size of $\pm 1$ Mpc. We have checked that the derivatives are stable over a range of adjacent step sizes. The azimuthal symmetry of the collision signatures and our chosen central position guarantee that the derivatives with respect to $\dphi$, $\cdxs$ and $\xc$ are non-zero for $m=0$ only.

Though the templates in general depend on the angular position of the centre of the bubble collision, for our particular choice of centre ($\angs = (0, 0)$) we can neglect both coordinates from the Fisher matrix analysis. Firstly, the derivative with respect to the longitude $\phi$ is zero due to the azimuthal symmetry of the signatures. Secondly, the derivative with respect to the colatitude $\theta$ --- formed by rotating the templates to $(\pm \delta \theta, 0)$ and calculating two-sided finite differences --- is non-zero for odd $m$ only. As a result, all off-diagonal entries in the Fisher matrix involving $\theta$ are zero, and $\theta$ is therefore entirely uncorrelated with the parameters of interest. The Fisher matrix we calculate therefore reduces to
\begin{equation}
\fisher_{ij} = \sum_{\ell} \frac{ \frac{\partial\blz^T}{\partial\theta_i} \frac{\partial\blz^T}{\partial\theta_j} \cell^{EE} - \left( \frac{\partial\blz^T}{\partial\theta_i} \frac{\partial\blz^E}{\partial\theta_j} + \frac{\partial\blz^E}{\partial\theta_i} \frac{\partial\blz^T}{\partial\theta_j} \right) \cell^{TE} + \frac{\partial\blz^E}{\partial\theta_i} \frac{\partial\blz^E}{\partial\theta_j} \cell^{TT}}{ \cell^{TT} \cell^{EE} - (\cell^{TE})^2 },
\label{eq:final_fisher}
\end{equation}
where we have absorbed the noise power spectra into $\cell^{XX}$.

We evaluate the Fisher matrix on a 50x50x13 grid of parameter values, with linearly spaced samples in the ranges $0.0005 \le \dphi \le 0.014$, $0.01\le \cdxs \le 2.0$ and $1000 \le \xc \le 13000$ Mpc. At each sampled point in parameter space we report the fractional marginalized uncertainty on each of the parameters of interest: the Lagrangian parameters $\dphi$, $\cdxs$ and $\xc$, and the observable amplitudes $\rzerolin$, $\rzeroquad$ and $\rzero$.

\subsection{Checks}

Although it is prohibitively slow to perform a brute-force likelihood analysis on this grid, we can evaluate the likelihood on a smaller grid to check our Fisher matrix outputs for the Lagrangian parameters. The likelihood we need to calculate is the probability of obtaining the data (which we assume contain a bubble, $\blm$, with some fiducial parameter values) assuming our model (that the data contain a bubble, $\tlm$, with a sampled set of parameters). To connect with the Fisher matrix analysis, we have to marginalize the logarithm of this likelihood over all CMB realizations. In the setting described above, this quantity takes the following form
\begin{align}
\langle \Delta \chi_{\rm bub}^2 \rangle \,\, = \sum_\ell & \left[ \frac{ \left[ (\blz^T)^2 - 2 \blz^T \tlz^T + (\tlz^T)^2 \right] \cell^{EE} + \left[ (\blz^E)^2 - 2 \blz^E \tlz^E + (\tlz^E)^2 \right] \cell^{TT} - }{ \cell^{TT} \cell^{EE} - (\cell^{TE})^2 } \right.\\
& \left. \frac{ 2 \left[ \blz^T \blz^E - \blz^T \tlz^E - \blz^E \tlz^T + \tlz^T \tlz^E \right] \cell^{TE} }{ \cell^{TT} \cell^{EE} - (\cell^{TE})^2 } \right],
\end{align}
where terms independent of the bubble collision parameters have been discarded.

We evaluate this expression on a 50x50 grid of $\dphi$ and $\cdxs$, selecting $\xc$ = 10 Gpc as a fiducial bubble size. The one-sigma {\em conditional} errors on each parameter are determined by finding the points at which the change in $\chi^2$ is one, linearly interpolating between grid points for extra accuracy. For $\dphi$, we find sub-percent-level agreement between the errors derived via the Fisher matrix and the brute-force likelihood evaluation: this is to be expected, as the likelihood is Gaussian in $\dphi$. For $\cdxs$, in which the likelihood is not Gaussian, we find the Fisher matrix overestimates the uncertainty by factors of a few percent (when the error contours are well resolved by the grid) to factors of $\sim 2$ (where interpolation errors are large). In the regime where interpolating errors are small, we conclude that the Lagrangian parameter constraints presented below are accurate at the percent level despite the non-Gaussian nature of the likelihood for $\cdxs$. The likelihood for $\rzerolin$ and $\rzeroquad$ is Gaussian, and therefore we expect the Fisher matrix constraints to reproduce results obtained from the exact likelihood.


\section{Results}\label{sec:results}

\begin{figure}
  \begin{center}
    \includegraphics[width=0.49\textwidth]{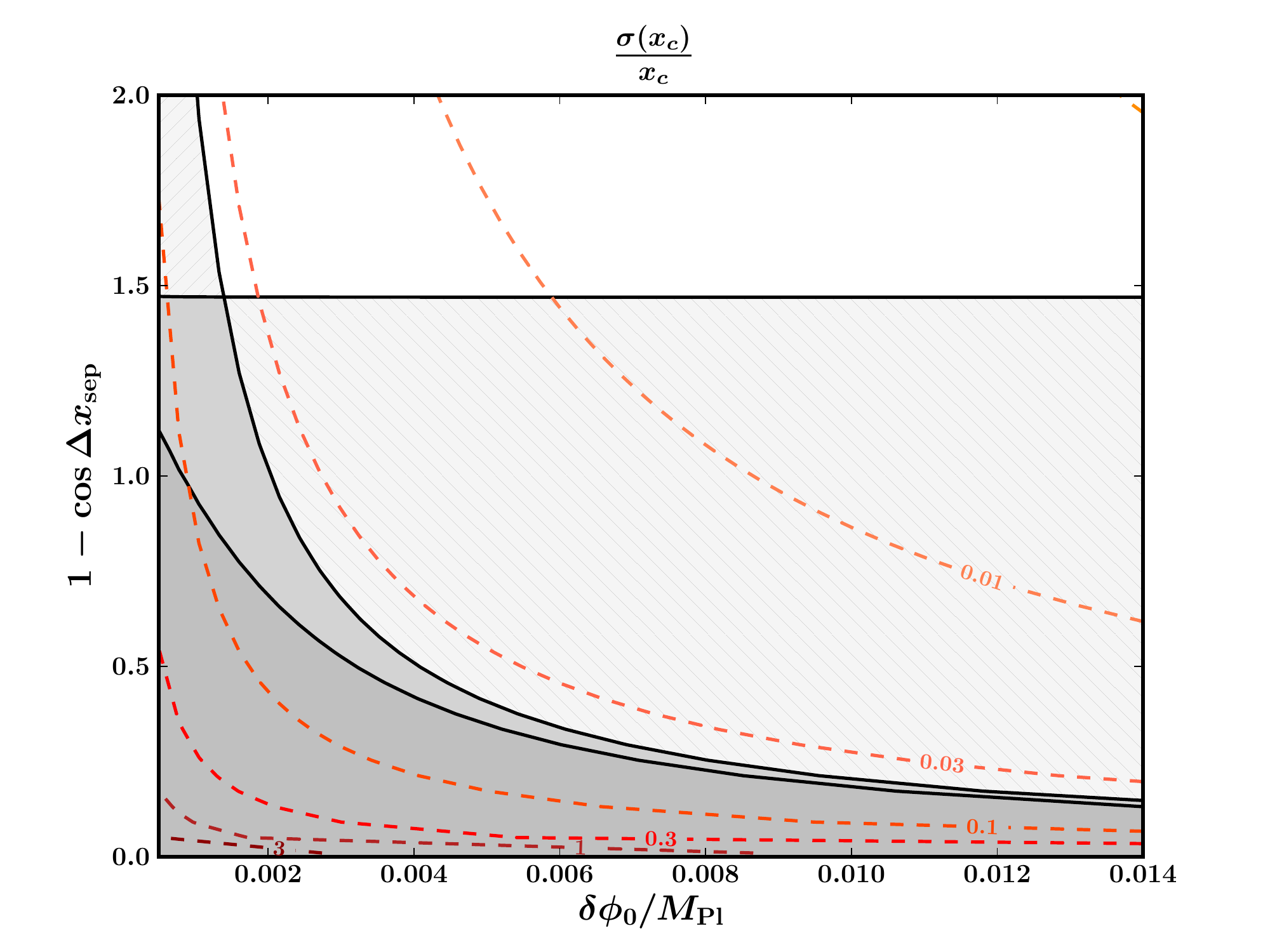}
    \includegraphics[width=0.49\textwidth]{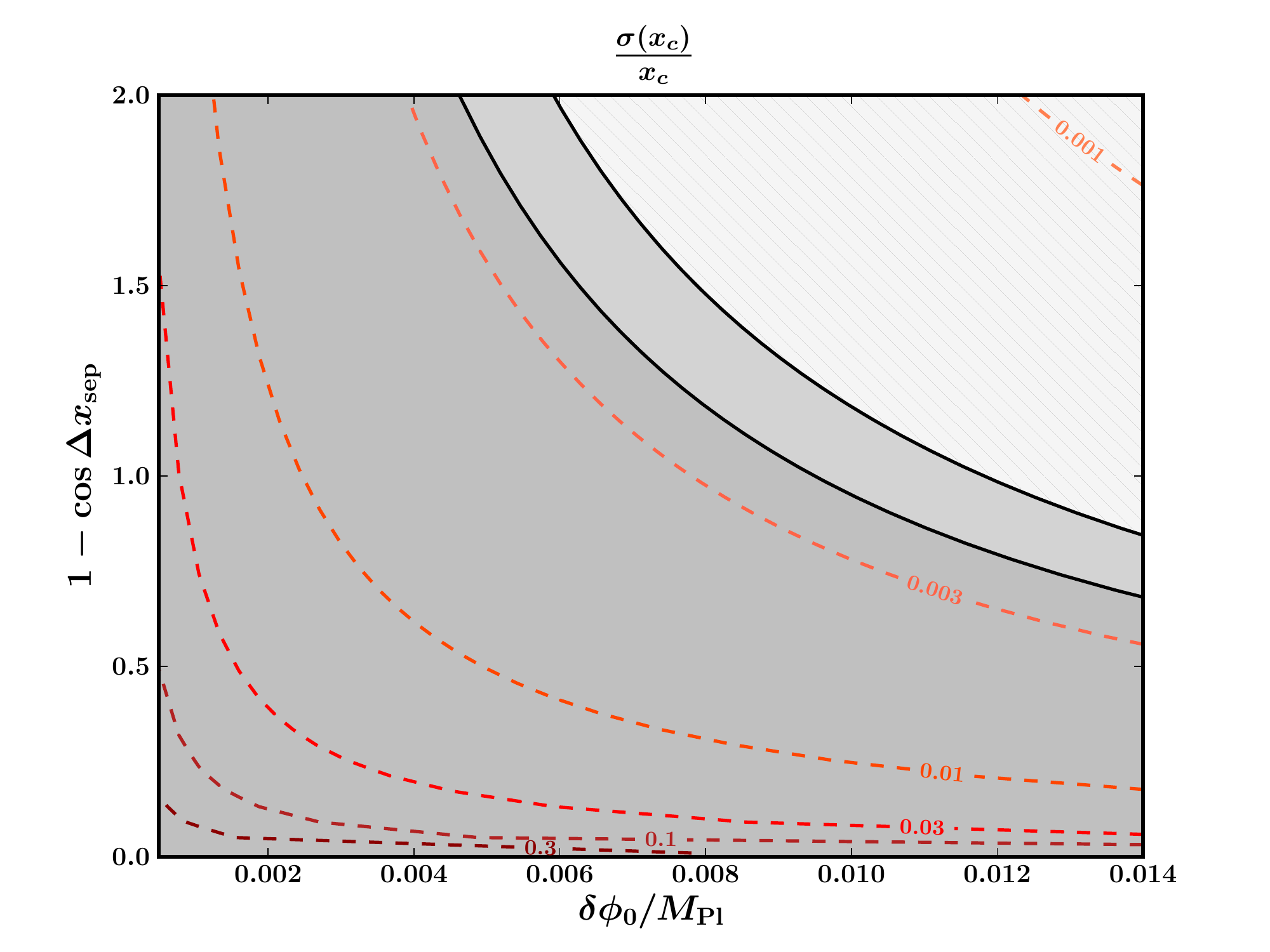}
    \includegraphics[width=0.49\textwidth]{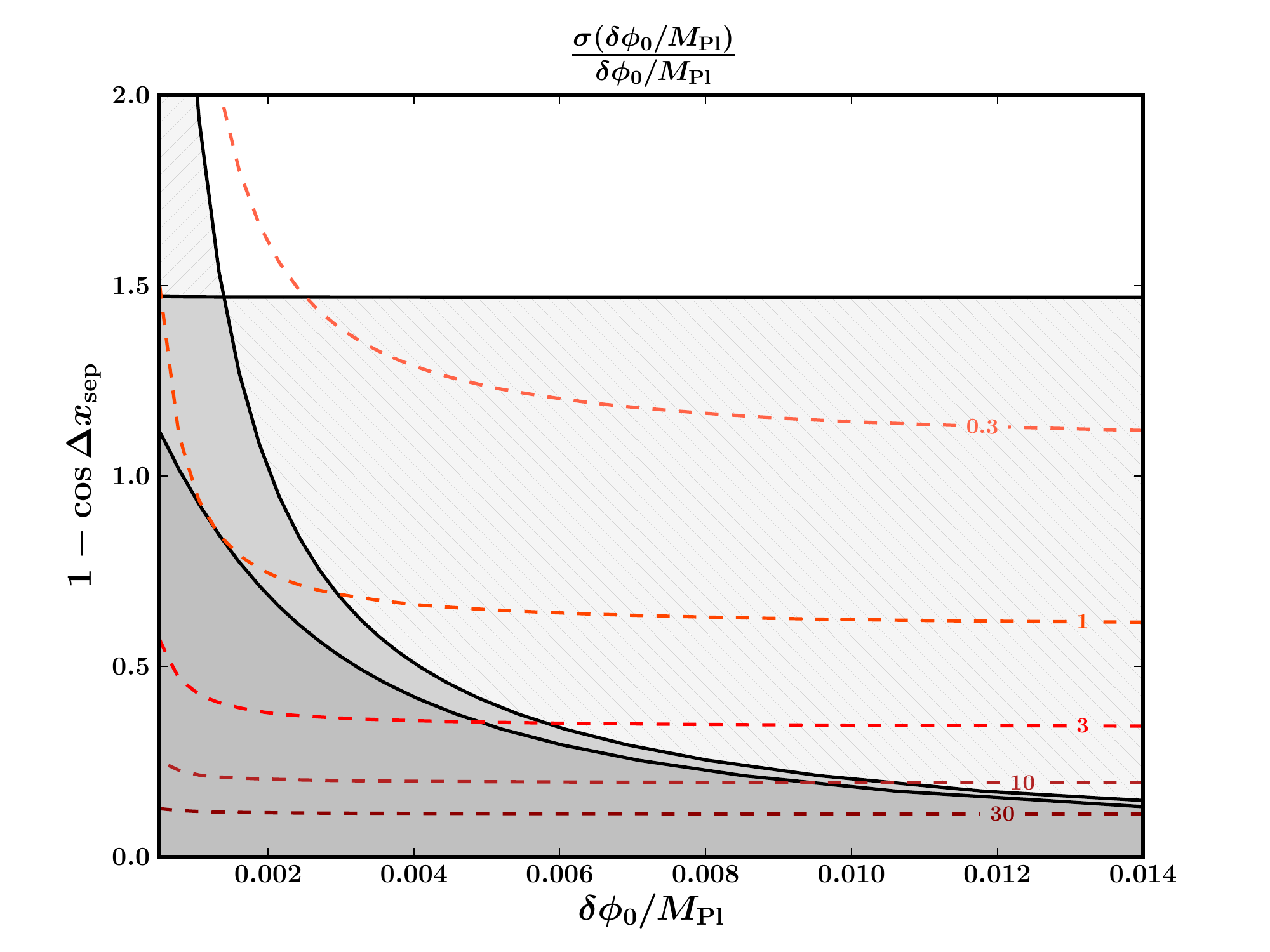}
    \includegraphics[width=0.49\textwidth]{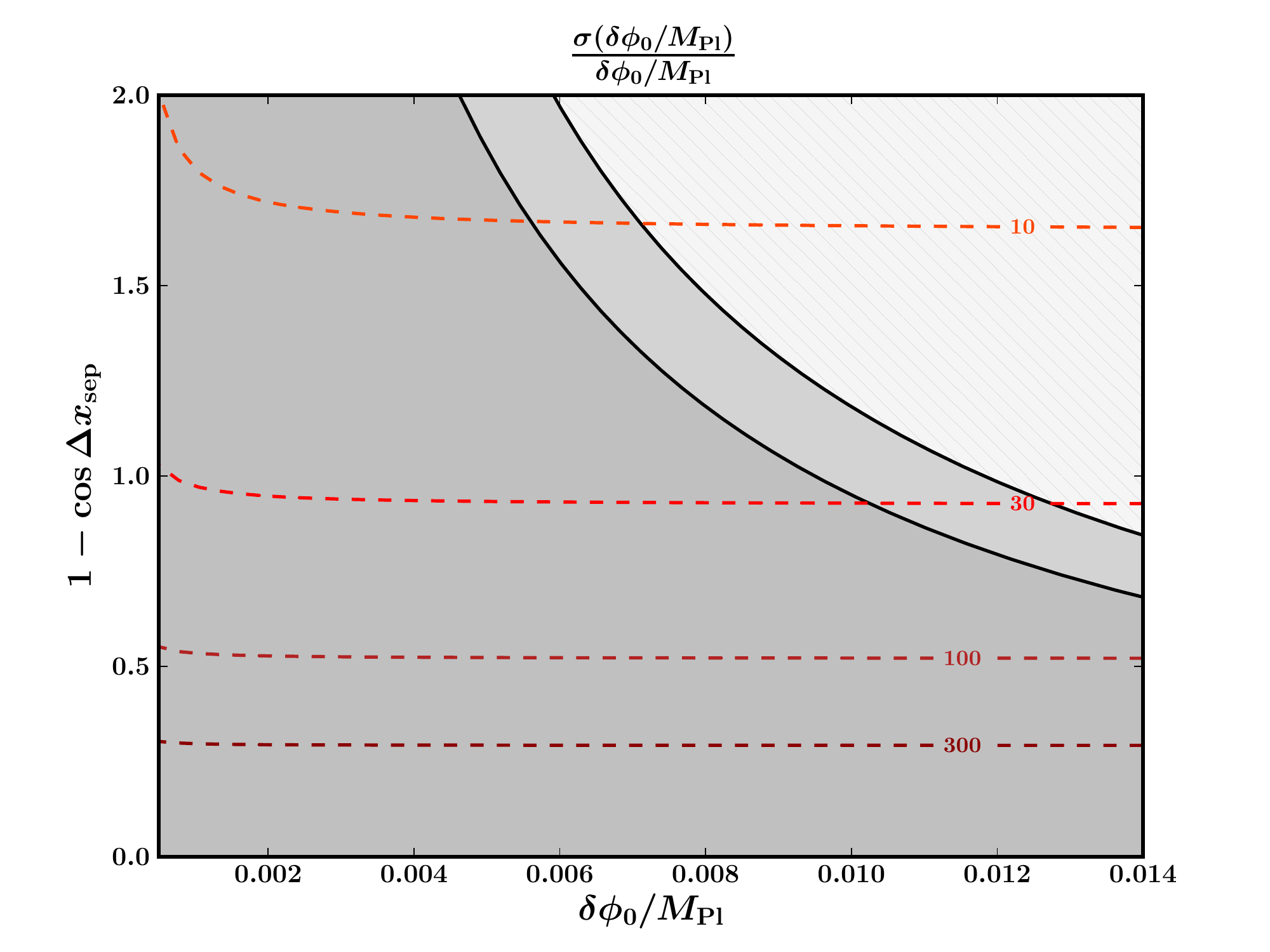}
    \includegraphics[width=0.49\textwidth]{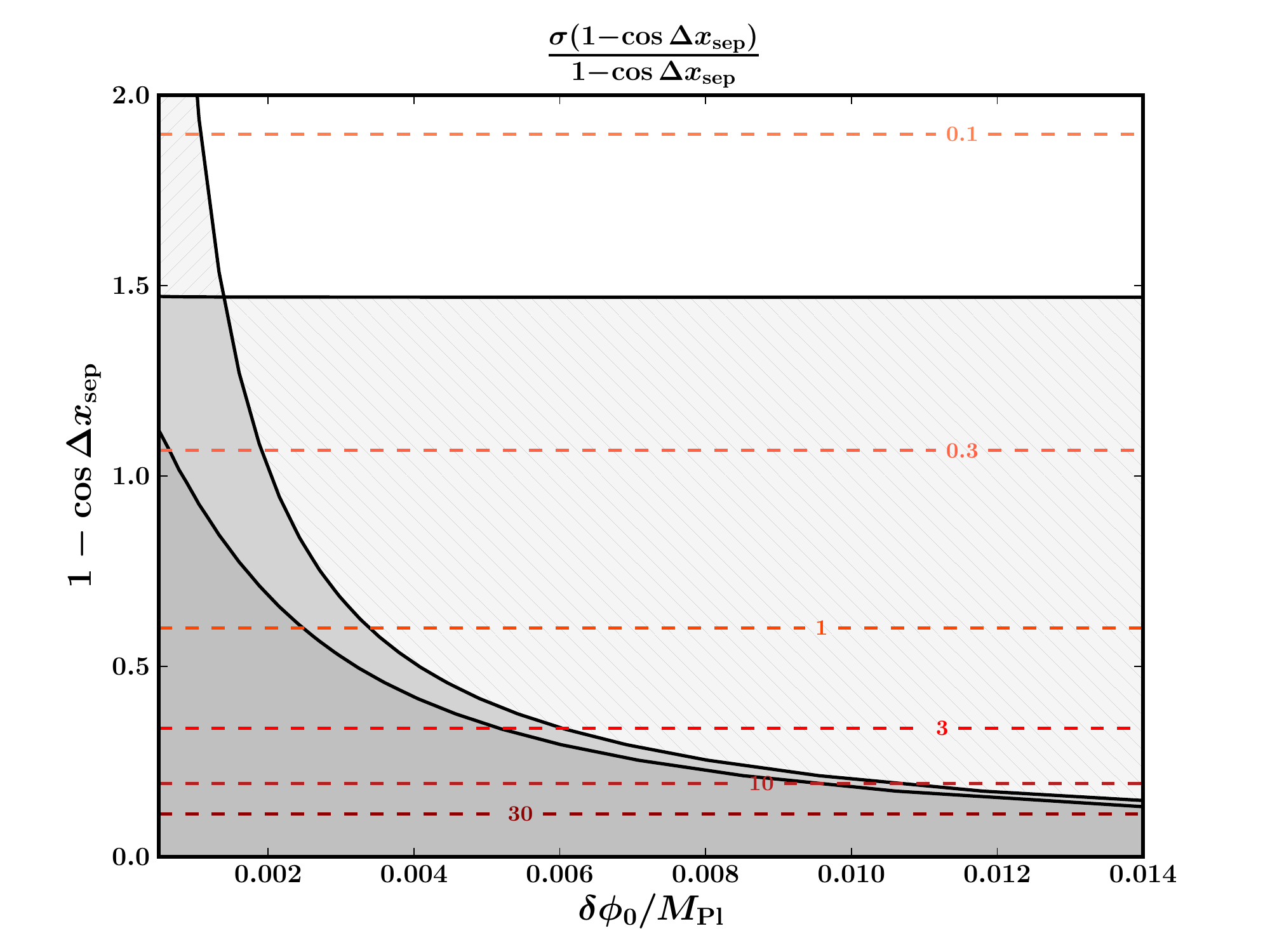}
    \includegraphics[width=0.49\textwidth]{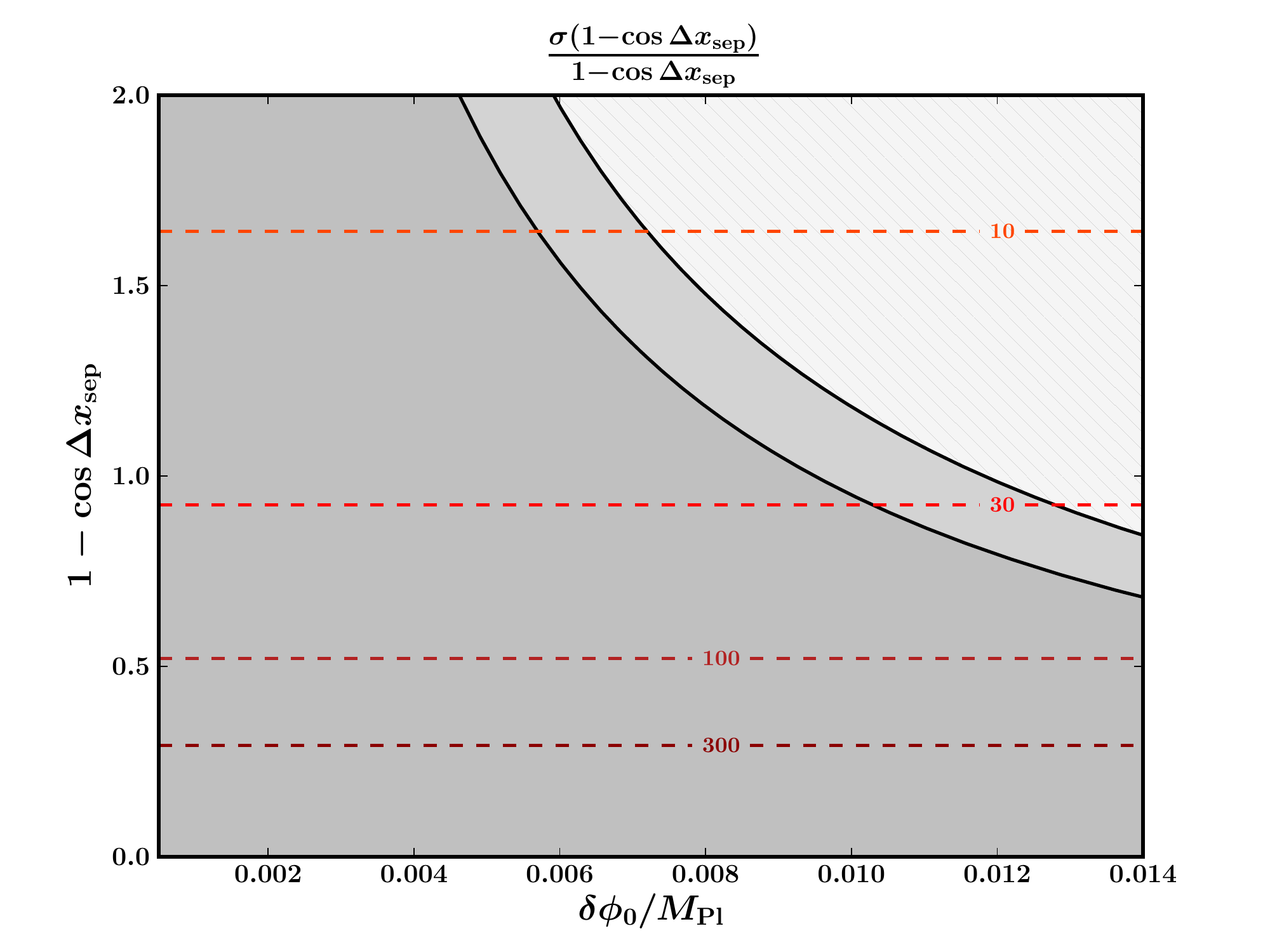}
    \caption{Contour plots of the fractional marginalized uncertainty on the Lagrangian parameters (\emph{dashed, red to orange}) for signatures with $\xc=1$ (\emph{left column}) and 13 Gpc (\emph{right column}). Regions of parameter space in which different components are distinguishable from zero at $3\sigma$ are indicated by shading. In dark grey regions, none of the amplitudes are detectable; in mid grey regions, $\rzero$ is detectable but $\rzerolin$ and $\rzeroquad$ are not; in light grey \hatchleft hatched regions, $\rzerolin$ is detectable in addition to $\rzero$; in light grey \hatchright hatched regions, $\rzeroquad$ is detectable in addition to $\rzero$. In white regions all amplitudes are detectable.}
    \label{fig:param_contours}
  \end{center}
\end{figure}

Illustrative examples of the parameter uncertainties achievable for the {\em cosmic-variance-limited} experiment are plotted in Fig.~\ref{fig:param_contours} for signatures with $\xc = 1$ Gpc (left) and $13$ Gpc (right). Each panel contains coloured contours of constant fractional uncertainty in a Lagrangian parameter. Superimposed are shaded regions in which certain sets of the amplitudes are detectable at the three-sigma level. More precisely:
\begin{enumerate}
\item in dark grey regions, none of the amplitudes are deemed detectable;
\item in mid-grey regions, the total amplitude ($\rzero$) is detectable, but the linear ($\rzerolin$) and quadratic ($\rzeroquad$) amplitudes are not;
\item in light grey \hatchleft hatched regions, the total and {\em linear} amplitudes are detectable;
\item in light grey \hatchright hatched regions, the total and {\em quadratic} amplitudes are detectable; and
\item in white regions, all amplitudes are detectable.
\end{enumerate}

A brief comment on theoretical priors is appropriate at this stage. Assuming that bubble nucleation is a random stochastic process yields predictions for the theoretical priors over $\xc$ and $\dxs$. The prior over $\xc$ is uniform~\cite{Freivogel_etal:2009it,Aguirre:2009ug}, and the prior over $\dxs$ is proportional to $\sin^3 \dxs$~\cite{Aguirre:2009ug} . Without a better microphysical theory of the potentials underlying eternal inflation, the prior over $\dphi$ is unknown; however, a reasonable assumption would be to employ a logarithmic prior. In the signal-dominated regime with which we are primarily concerned, these priors do not significantly change the conclusions presented below.

\subsection{Template Detectability}

In general, the detectability boundaries (solid black contours in Fig.~\ref{fig:param_contours}) follow lines of constant amplitude ($\rzero$, $\rzerolin$, or $\rzeroquad$). Note that the detectability boundary for the quadratic amplitude is missing from the $\xc = 13$ Gpc plot (and, indeed, for all $\xc \gtrsim 5$ Gpc): there is no combination of the Lagrangian parameters at these scales for which a pure-quadratic bubble collision signature is detectable. The reason for this is clear from the definition of the quadratic amplitude (Eq.~\ref{eq:rzeroquad_trans}). As $\xc$ tends to $\xls$, the $\xcnorminv$ term completely overwhelms the $\cdxs$ term, strongly suppressing $\rzeroquad$. Where it is present, the detectability boundary for the quadratic amplitude is, as expected, a function of $\cdxs$ only: for a given scale, $\rzeroquad$ depends only on $\cdxs$. This relation breaks down for very small values of $\dphi$ (a factor of  $\sim10$ smaller than the range plotted in Fig.~\ref{fig:param_contours}), at which point a degeneracy opens up between $\rzeroquad$ and $\xc$. Here, the detectability boundary turns upwards, becoming a strong function of $\dphi$. The sharper features of the linear template allow $\xc$ to be better determined.

Though $\rzerolin$ also contains a factor of $\xcnorminv$ (Eq.~\ref{eq:rzerolin_trans}), we can always find a value of $\dphi$ large enough to make it detectable. Unlike the quadratic amplitude, the linear amplitude is therefore detectable in some portion of the $\dphi$--$(\cdxs)$ plane for all values of $\xc$ considered, albeit at higher and higher $\dphi$ as $\xc$ tends to $\xls$. As the product of $\dphi$ and $\cdxs$ appears in the linear amplitude (Eq.~\ref{eq:rzerolin_trans}) and $\xc$ is well determined in the parameter range considered here, the $\rzerolin$ detectability boundary is given by $(\cdxs) \dphi \propto$ const. 

To a first approximation, the total amplitude detectability boundary can be thought of as interpolating between the pure-linear and pure-quadratic cases. In fact, the linear and total amplitude boundaries cross over for very small $\rzeroquad$. In this regime, examination of Eq.~\ref{eq:template_obs_tot} shows that we should expect $\rzero$ and $f$ to become completely degenerate, as we are trying to fix both of their values using only one number ($\rzerolin$). The $\rzero, f$ parameterization is not a good model for regimes in which one of the templates is negligible.

\subsection{Lagrangian Parameter Constraints}

We now turn our focus to the Lagrangian parameter constraints (the coloured contours in Fig.~\ref{fig:param_contours}). Recall that these constraints depend on the detection of spatial curvature and primordial tensors, where we have assumed $\ro = 0.1$ and $\omk = 10^{-4}$. 

First, we note that $\xc$ (top) is the best constrained of the Lagrangian parameters, with percent-level errors achievable across much of the parameter space considered. Note that the shapes of the constant-$\xc$-error contours mirror that of the linear amplitude decision boundary: the sharp features in the linear template are critical in determining the size of the signature.

The performance for the other two Lagrangian parameters, $\dphi$ and $\cdxs$, is much poorer, and highly dependent on the size of signature present. We only obtain direct constraints on $\cdxs$ through the quadratic template, with the linear template constraining the product $\dphi \left[ \cdxs \right]$. This correlation structure is evident in the shapes of the individual contours. For signatures covering a large fraction of the sky we are able to measure both the linear and quadratic amplitudes, and hence constrain $\dphi$ and $\cdxs$ with accuracies of up to $\sim10\%$. The smallest values of $\delta\phi_0^{\rm coll}$ and $\dxs$ measurable at three-sigma are roughly $M_{\rm Pl}/500$ and $\pi/2$, respectively. For smaller signatures, our inability to accurately measure the amplitude of the quadratic template, compounded by the the factor of $\xcnorminv$ appearing in the linear amplitude, means the uncertainties in both parameters grow to the point where order-of-magnitude estimates are no longer possible, even with a full-sky, infinite-resolution, cosmic-variance-limited experiment.

\subsection{Comparing Datasets and Experiments}

\begin{figure}
  \begin{center}
    \includegraphics[width=1.0\textwidth]{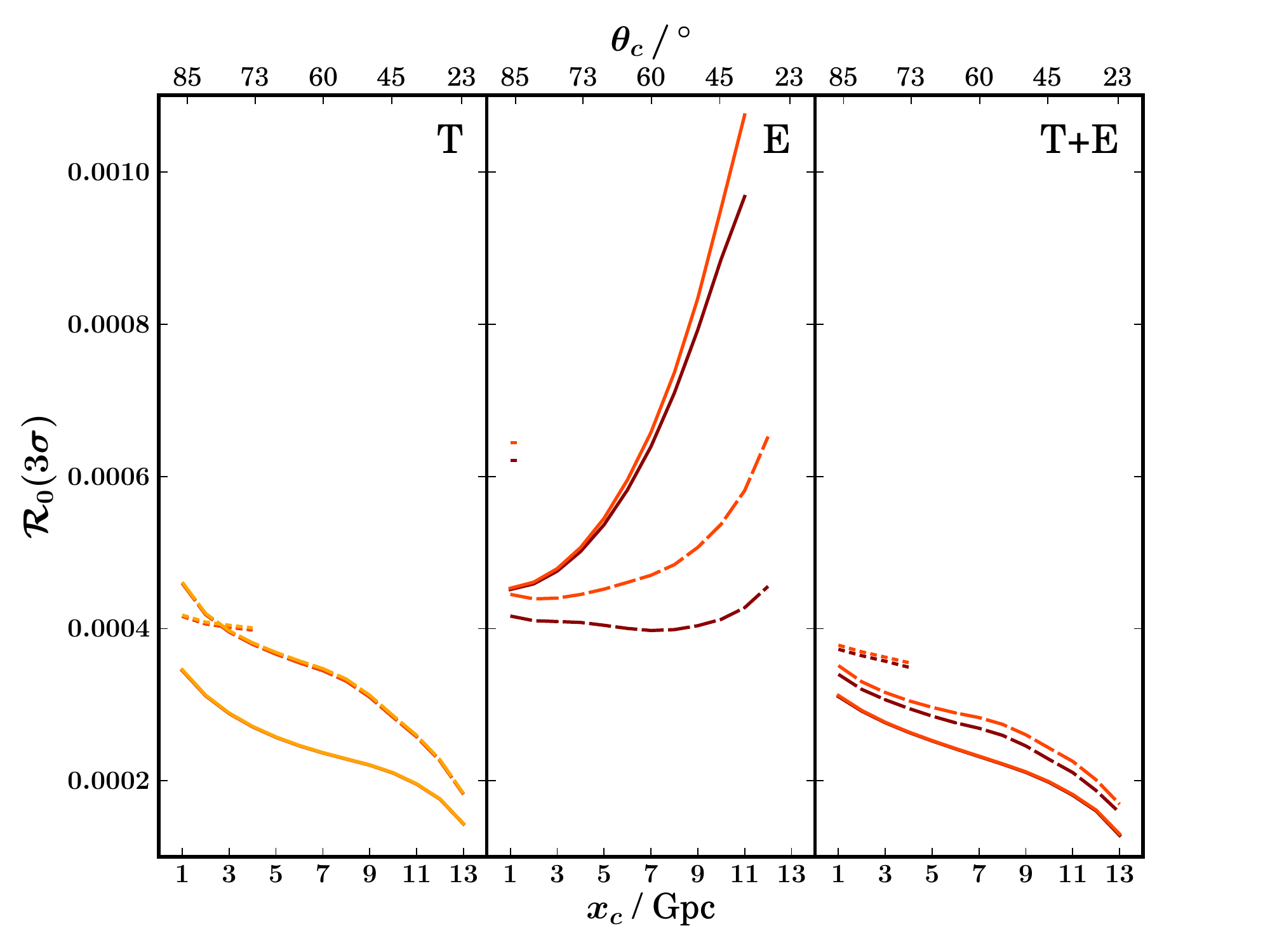}
    \caption{Amplitudes detectable at $3\sigma$ using temperature-only (\emph{left}), polarization-only (\emph{center}) and combined temperature and polarization (\emph{right}) data from WMAP7 ({\em orange}), \Planck\ (\emph{red}) and a cosmic-variance-limited experiment (\emph{dark red}). Detectable values of $\rzero$ are indicated with solid lines, $\rzerolin$ with long-dashed lines and $\rzeroquad$ with short-dashed lines.}
    \label{fig:3_sig_amps}
  \end{center}
\end{figure}

As discussed above, the detectability boundaries correspond to constant-amplitude curves in the $\dphi$--$(\cdxs)$ plane. These amplitudes are plotted as a function of $\xc$ in Fig.~\ref{fig:3_sig_amps} for all experiments considered. Fig.~\ref{fig:3_sig_amps} contains three panels, displaying the amplitudes detectable using temperature-only, polarization-only and combined temperature and polarization information.

\begin{figure}
  \begin{center}
    \includegraphics[width=.5\textwidth]{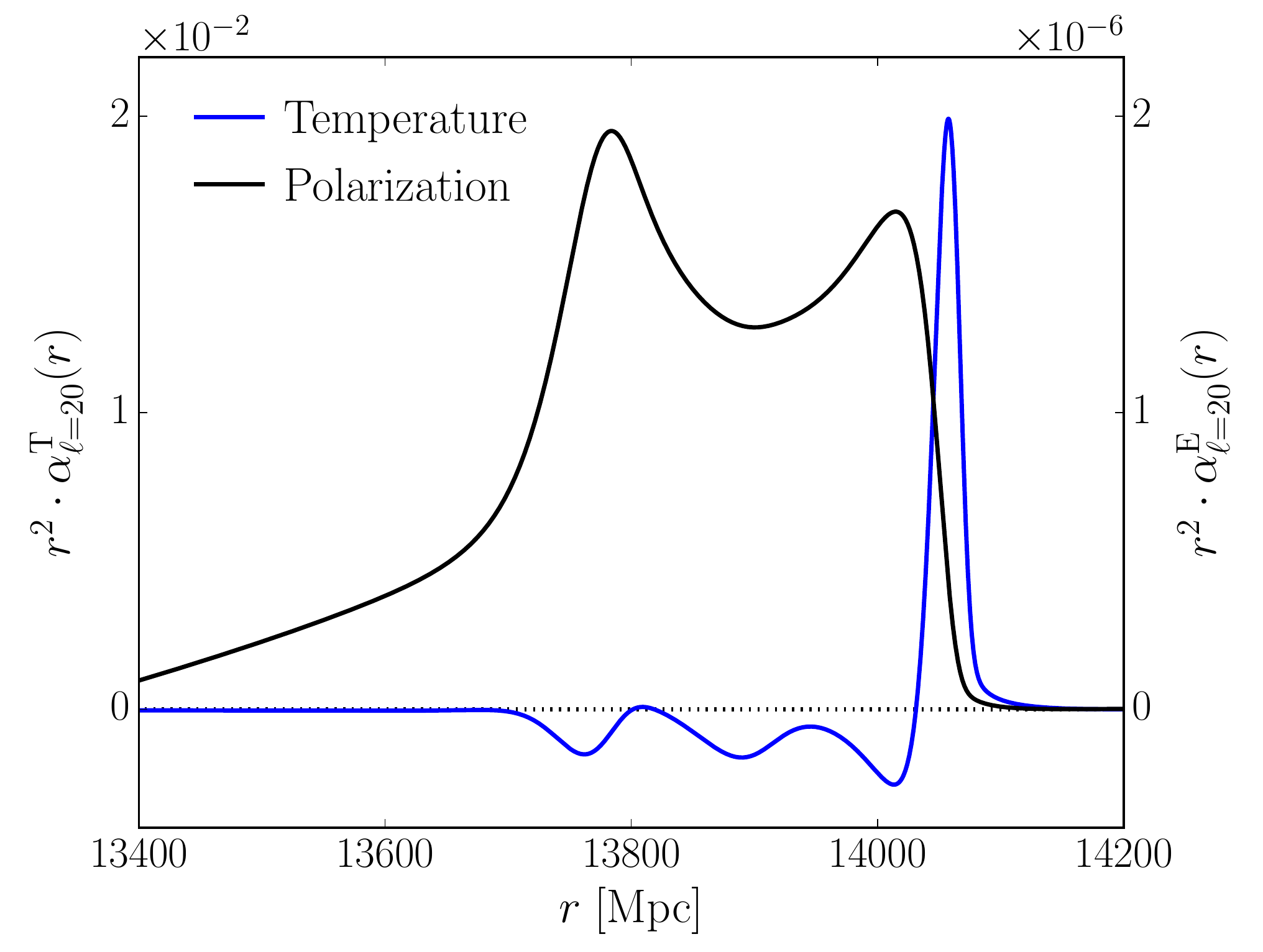}
    \caption{At large scales, the temperature and polarization
      transfer functions behave qualitatively differently. While for
      polarization, they remain positive over the full width of the
      last scattering surface (\emph{black solid line, scale shown on
        the right y-axis}), for temperature, a change of sign occurs,
      leading to partial cancellation of the curvature perturbation
      contribution to the template signal (\emph{blue solid line,
        scale shown on the left y-axis}).}
    \label{fig:te_transferfunctions}
  \end{center}
\end{figure}

Concentrating first on the overall trends, we note that when only temperature information is used the detectable amplitudes decrease with $\xc$, but when only polarization information is used the detectable amplitudes in general increase with $\xc$. This is because the primordial curvature templates are normalized to the same central value, but the same is not true for the evolved CMB templates: there is slight variation in the central amplitude of the CMB templates ($\normlinlm^X, \normquadlm^X$) with $\xc$. In particular, the maximum values of $\normlinlm^T$ and $\normquadlm^T$ increase with $\xc$, whereas the maximum values of $\normlinlm^E$ and $\normquadlm^E$ decrease with $\xc$.
This behavior can be explained from the shape of the real space
transfer functions used in the template computation
(Eq.~\ref{eq:r2alm}) on large scales. In Fig.~\ref{fig:te_transferfunctions}, we
display the functional form of the integrand $r^2 \alpha_{\ell}(r)$
for the multipole moment $\ell = 20$, corresponding to angular
scales of $\sim 10^{\circ}$. While the transfer function is
larger than zero over the full width of the last scattering surface in
polarization, in temperature, it changes sign from being positive at
the location of the peak of the photon visibility function to being
negative at smaller radii. For strictly non-negative curvature
perturbations, contributions to the CMB temperature template will
therefore partially cancel out, generating larger template amplitudes
for more quickly decaying perturbations (i.e., larger $\xc$). For
polarization, on the other hand, no such cancellation occurs and the
CMB template amplitude increases for decreasing $\xc$.

Turning to $\rzeroquad$, we again note that the quadratic signature is only detectable for a small range of the $\xc$ values considered. The detectable $\rzeroquad$ is smallest (and the range of $\xc$ over which $\rzeroquad$ is detectable is largest) when combining temperature and polarization. The CMB temperature provides stronger constraints than the polarization. Considering $\rzerolin$ instead, the sharp polarization feature present at $\thetac$ in the linear template means that $\rzerolin$, in contrast to $\rzeroquad$, is well constrained by polarization alone; indeed it is better constrained by polarization in the range $\xc \lesssim 3$ Gpc. Temperature is, however, much better for the total amplitude $\rzero$ (red), as the quadratic template is so poorly constrained by polarization.

For the overall amplitude, $\rzero$, the detectability limits are mostly determined by temperature data. Note that the linear amplitude is always better constrained than the overall amplitude when employing only polarization, but the opposite is true when using temperature-only data. This is because the amplitude of the linear polarization template (specifically, the sharp feature at $\thetac$) is around twice that of the quadratic polarization template, whereas the temperature templates are very similar in amplitude ($\normquadlm^T$ is typically $\sim 10 \%$ stronger than $\normlinlm^T$). The $\rzero, f$ degeneracy is therefore exposed at larger $\cdxs$ when using polarization than when using only temperature. When temperature and polarization data are combined, the detectable $\rzero$ and $\rzerolin$ curves are very similar.

Comparing experiments, the WMAP, \Planck\  and cosmic-variance-limited curves all converge when using temperature-only information: both WMAP and \Planck\ are essentially cosmic-variance-limited in temperature at the multipoles relevant to this analysis. Indeed, \Planck\ is also nearly cosmic-variance-limited in E polarization, yielding constraints on $\rzero$ almost indistinguishable from the cosmic-variance-limited experiment, and constraints on $\rzerolin$ only $7\%$ weaker at the largest angular scales. Though \Planck's polarization noise becomes more important for smaller collision signatures, its combined temperature and polarization data yield detectable amplitudes a maximum of only $8\%$ larger than the cosmic variance limit on the smallest scales considered here. As the constraints on $\rzero$ are temperature-dominated, \Planck\ offers less than a $10\%$ improvement over WMAP; however, its polarization data allow up to $\sim 30\%$ improvement in constraints on the linear amplitude, $\rzerolin$, for collision signatures covering half the sky. Beyond increasing the fraction of foreground-cleaned sky available for analysis, there is little improvement to be gained from post-\Planck\ data.


\section{Comparing with previous work}\label{sec:comparison}

The forecasts presented above are broadly consistent with previous work on constraining bubble collisions using data from the WMAP satellite. Two groups have presented constraints on the linear portion of the template. The most recent analysis of Feeney et. al.~\cite{Feeney:2012hj} implemented a hierarchical Bayesian formalism which tests the bubble collision model on the full sky, constraining the expected total number of collisions to be fewer than 4 at $95\%$ confidence. To facilitate comparison with the single-template forecast presented above, we consider their candidate-detection step which implements an optimal filtering of the CMB for templates of different scales. The threshold for $50 \%$ of bubble collisions to be detected with a signal-to-noise ratio of greater than 3 is $\Delta T / T (\angs) > 6.5 \times 10^{-5}$, and is relatively independent of the angular scale of the collision $\thetac$. Neglecting the evolution of the comoving curvature perturbation, in the Sachs-Wolfe approximation, where $\Delta T/T \simeq \mathcal{R}/5$, this implies $\rzerolin > 3.3 \times 10^{-4}$. We take this to be the rough constraint on individual templates implied by the results of Feeney et. al. Osborne et. al.~\cite{Osborne:2013hea} implemented an optimal estimator for single collision templates, directly constraining $\rzerolin$ to $|\rzerolin| < 6.7 \times 10^{-4} (1 - \cos \thetac) (\sin \thetac)^{-4/3}$ at $2 \sigma$.\footnote{This is obtained by relating the parameter $a$ in Ref.~\cite{Osborne:2013hea} to $\rzerolin$ by $\rzerolin = a (\xls - \xc)$.} Both of these constraints are compatible with the WMAP forecast presented in Fig.~\ref{fig:3_sig_amps}; however the presence of sky cuts and other complications lead to somewhat weaker constraints from WMAP data than the full-sky forecast presented above.

A competing method for detecting bubble collisions is to use the technique of kinetic Sunyaev-Zel'dovich tomography~\cite{Zhang:2015uta}. This method circumvents cosmic variance by looking at the effect of a collision on the velocity field of free electrons, measured using the cross-correlation between large scale structure and the kinetic Sunyaev-Zel'dovich effect in the CMB. A combination of Large Synoptic Survey Telescope (LSST)~\cite{2009arXiv0912.0201L} and \Planck\ temperature data is forecasted to yield constraints of $\rzerolin < 2 \times 10^{-4}$ and $\rzeroquad < 7.5 \times 10^{-4}$ at $3 \sigma$ for collisions on the largest angular scales. These constraints can be improved by an order of magnitude by a future high-resolution CMB experiment, with an ultimate cosmic-variance limit of $\rzerolin, \rzeroquad \lesssim 10^{-9}$. 


\section{Discussion and conclusions}\label{sec:discussion}

We have forecasted the ability of cosmic-variance-limited CMB datasets, employing information from both temperature and polarization, to constrain theories of eternal inflation using bubble collisions. We have studied both the raw detectability for bubble collisions, as well as the ability of CMB datasets to constrain fundamental parameters of the theory underlying eternal inflation. Comparing the detectability achievable with a cosmic-variance-limited CMB experiment and data from the \Planck\ satellite to existing constraints arising from WMAP temperature data, we find that temperature data yields virtually no improvement on current limits. Adding polarization data improves limits on detectability by $\lesssim 30\%$, and there are only marginal improvements between \Planck\ and the cosmic-variance-limited experiment. We can therefore view data from \Planck\ as having the final word on bubble collisions in the CMB. 

The bubble collision template for the comoving curvature perturbation in Eq.~\ref{eq:raw_template}, once evolved into the CMB temperature and polarization anisotropies, directly links the scalar field theory underlying eternal inflation to cosmological observables. In this paper, we have restricted ourselves to collisions between identical bubbles, in which case the observables are determined entirely by the width of the barrier in the scalar-field potential (fixed for a given theory) as well as the initial separation between bubbles and the observer position (stochastic variables). Assuming the detection of negative spatial curvature and primordial gravitational waves, which enter into the amplitudes for different components of the template, we have presented limits on detecting the barrier width, initial separation, and  observer position. In the optimistic scenario where negative spatial curvature and primordial gravitational waves are detected, we expect these to be close to the bounds achievable by an analysis of \Planck\ data.

Looking beyond the CMB, it is possible to use measurements of the kinetic Sunyaev-Zel'dovich effect to mine cosmological data further for signatures of bubble collisions. From the theoretical side, more-varied phenomenology could arise in theories of eternal inflation with multiple scalar fields, or for collisions between different bubbles. For example, recent work~\cite{Braden:2014cra,Braden:2015vza,Bond:2015zfa} has shown a variety of models will break the assumed symmetry of the collision spacetime, possibly leading to a collision signature with more structure. We leave an exhaustive study of constraints on other models to future work.


\acknowledgments

SMF thanks Alan Heavens and Andrew Jaffe for numerous illuminating discussions. SMF is supported by the Science and Technology Facilities Council in the UK. FE and HVP are supported by the European Research Council under the European Community's Seventh Framework Programme (FP7/2007-2013) / ERC grant agreement no 306478-CosmicDawn. Research at Perimeter Institute is supported by the Government of Canada through Industry Canada and by the Province of Ontario through the Ministry of Research and Innovation. MCJ is supported by the National Science and Engineering Research Council through a Discovery grant. 


\bibliography{cv_te_forecast.bib}

\begin{thebibliography}{60}
\expandafter\ifx\csname natexlab\endcsname\relax\def\natexlab#1{#1}\fi
\expandafter\ifx\csname bibnamefont\endcsname\relax
  \def\bibnamefont#1{#1}\fi
\expandafter\ifx\csname bibfnamefont\endcsname\relax
  \def\bibfnamefont#1{#1}\fi
\expandafter\ifx\csname citenamefont\endcsname\relax
  \def\citenamefont#1{#1}\fi
\expandafter\ifx\csname url\endcsname\relax
  \def\url#1{\texttt{#1}}\fi
\expandafter\ifx\csname urlprefix\endcsname\relax\def\urlprefix{URL }\fi
\providecommand{\bibinfo}[2]{#2}
\providecommand{\eprint}[2][]{\url{#2}}

\bibitem[{\citenamefont{{Tauber} et~al.}(2010)\citenamefont{{Tauber},
  {Mandolesi}, {Puget} et~al.}}]{Tauber2010}
\bibinfo{author}{\bibfnamefont{J.~A.} \bibnamefont{{Tauber}}},
  \bibinfo{author}{\bibfnamefont{N.}~\bibnamefont{{Mandolesi}}},
  \bibinfo{author}{\bibfnamefont{J.}~\bibnamefont{{Puget}}},
  \bibnamefont{et~al.}, \bibinfo{journal}{A\&A} \textbf{\bibinfo{volume}{520}}
  (\bibinfo{year}{2010}).

\bibitem[{\citenamefont{{Hawking}}(1982)}]{1982PhLB..115..295H}
\bibinfo{author}{\bibfnamefont{S.~W.} \bibnamefont{{Hawking}}},
  \bibinfo{journal}{Physics Letters B} \textbf{\bibinfo{volume}{115}},
  \bibinfo{pages}{295} (\bibinfo{year}{1982}).

\bibitem[{\citenamefont{{Starobinsky}}(1982)}]{1982PhLB..117..175S}
\bibinfo{author}{\bibfnamefont{A.~A.} \bibnamefont{{Starobinsky}}},
  \bibinfo{journal}{Physics Letters B} \textbf{\bibinfo{volume}{117}},
  \bibinfo{pages}{175} (\bibinfo{year}{1982}).

\bibitem[{\citenamefont{{Guth} and {Pi}}(1982)}]{1982PhRvL..49.1110G}
\bibinfo{author}{\bibfnamefont{A.~H.} \bibnamefont{{Guth}}} \bibnamefont{and}
  \bibinfo{author}{\bibfnamefont{S.-Y.} \bibnamefont{{Pi}}},
  \bibinfo{journal}{Physical Review Letters} \textbf{\bibinfo{volume}{49}},
  \bibinfo{pages}{1110} (\bibinfo{year}{1982}).

\bibitem[{\citenamefont{{Bardeen} et~al.}(1983)\citenamefont{{Bardeen},
  {Steinhardt}, and {Turner}}}]{1983PhRvD..28..679B}
\bibinfo{author}{\bibfnamefont{J.~M.} \bibnamefont{{Bardeen}}},
  \bibinfo{author}{\bibfnamefont{P.~J.} \bibnamefont{{Steinhardt}}},
  \bibnamefont{and} \bibinfo{author}{\bibfnamefont{M.~S.}
  \bibnamefont{{Turner}}}, \bibinfo{journal}{Phys.Rev.}
  \textbf{\bibinfo{volume}{D28}}, \bibinfo{pages}{679} (\bibinfo{year}{1983}).

\bibitem[{\citenamefont{{Steinhardt}}(1983)}]{steinhardt1982}
\bibinfo{author}{\bibfnamefont{P.~J.} \bibnamefont{{Steinhardt}}}, in
  \emph{\bibinfo{booktitle}{{The very early universe. Proceedings of the
  Nuffield Workshop, held at Cambridge, England, 21 June - 9 July, 1982.}}}
  (\bibinfo{year}{1983}).

\bibitem[{\citenamefont{{Vilenkin}}(1983)}]{1983PhRvD..27.2848V}
\bibinfo{author}{\bibfnamefont{A.}~\bibnamefont{{Vilenkin}}},
  \bibinfo{journal}{Phys.Rev.} \textbf{\bibinfo{volume}{D27}},
  \bibinfo{pages}{2848} (\bibinfo{year}{1983}).

\bibitem[{\citenamefont{Aguirre et~al.}(2007)\citenamefont{Aguirre, Johnson,
  and Shomer}}]{Aguirre:2007an}
\bibinfo{author}{\bibfnamefont{A.}~\bibnamefont{Aguirre}},
  \bibinfo{author}{\bibfnamefont{M.~C.} \bibnamefont{Johnson}},
  \bibnamefont{and} \bibinfo{author}{\bibfnamefont{A.}~\bibnamefont{Shomer}},
  \bibinfo{journal}{Phys. Rev.} \textbf{\bibinfo{volume}{D76}},
  \bibinfo{pages}{063509} (\bibinfo{year}{2007}).

\bibitem[{\citenamefont{Hawking and Moss}(1982)}]{Hawking:1981fz}
\bibinfo{author}{\bibfnamefont{S.~W.} \bibnamefont{Hawking}} \bibnamefont{and}
  \bibinfo{author}{\bibfnamefont{I.~G.} \bibnamefont{Moss}},
  \bibinfo{journal}{Phys. Lett.} \textbf{\bibinfo{volume}{B110}},
  \bibinfo{pages}{35} (\bibinfo{year}{1982}).

\bibitem[{\citenamefont{Hawking et~al.}(1982)\citenamefont{Hawking, Moss, and
  Stewart}}]{Hawking:1982ga}
\bibinfo{author}{\bibfnamefont{S.~W.} \bibnamefont{Hawking}},
  \bibinfo{author}{\bibfnamefont{I.~G.} \bibnamefont{Moss}}, \bibnamefont{and}
  \bibinfo{author}{\bibfnamefont{J.~M.} \bibnamefont{Stewart}},
  \bibinfo{journal}{Phys. Rev.} \textbf{\bibinfo{volume}{D26}},
  \bibinfo{pages}{2681} (\bibinfo{year}{1982}).

\bibitem[{\citenamefont{Wu}(1983)}]{Wu:1984eda}
\bibinfo{author}{\bibfnamefont{Z.-C.} \bibnamefont{Wu}},
  \bibinfo{journal}{Phys. Rev.} \textbf{\bibinfo{volume}{D28}},
  \bibinfo{pages}{1898} (\bibinfo{year}{1983}).

\bibitem[{\citenamefont{Moss}(1994)}]{Moss:1994pi}
\bibinfo{author}{\bibfnamefont{I.~G.} \bibnamefont{Moss}}
  (\bibinfo{year}{1994}), \eprint{gr-qc/9405045}.

\bibitem[{\citenamefont{Aguirre and Johnson}(2008)}]{Aguirre:2007wm}
\bibinfo{author}{\bibfnamefont{A.}~\bibnamefont{Aguirre}} \bibnamefont{and}
  \bibinfo{author}{\bibfnamefont{M.~C.} \bibnamefont{Johnson}},
  \bibinfo{journal}{Phys. Rev.} \textbf{\bibinfo{volume}{D77}},
  \bibinfo{pages}{123536} (\bibinfo{year}{2008}).

\bibitem[{\citenamefont{Aguirre et~al.}(2009)\citenamefont{Aguirre, Johnson,
  and Tysanner}}]{Aguirre:2008wy}
\bibinfo{author}{\bibfnamefont{A.}~\bibnamefont{Aguirre}},
  \bibinfo{author}{\bibfnamefont{M.~C.} \bibnamefont{Johnson}},
  \bibnamefont{and} \bibinfo{author}{\bibfnamefont{M.}~\bibnamefont{Tysanner}},
  \bibinfo{journal}{Phys.Rev.} \textbf{\bibinfo{volume}{D79}},
  \bibinfo{pages}{123514} (\bibinfo{year}{2009}), \eprint{0811.0866}.

\bibitem[{\citenamefont{Aguirre and Johnson}(2011)}]{Aguirre:2009ug}
\bibinfo{author}{\bibfnamefont{A.}~\bibnamefont{Aguirre}} \bibnamefont{and}
  \bibinfo{author}{\bibfnamefont{M.~C.} \bibnamefont{Johnson}},
  \bibinfo{journal}{Rept.Prog.Phys.} \textbf{\bibinfo{volume}{74}},
  \bibinfo{pages}{074901} (\bibinfo{year}{2011}), \eprint{0908.4105}.

\bibitem[{\citenamefont{Kozaczuk and Aguirre}(2012)}]{Kozaczuk:2012sx}
\bibinfo{author}{\bibfnamefont{J.}~\bibnamefont{Kozaczuk}} \bibnamefont{and}
  \bibinfo{author}{\bibfnamefont{A.}~\bibnamefont{Aguirre}}
  (\bibinfo{year}{2012}), \eprint{1206.5038}.

\bibitem[{\citenamefont{Chang et~al.}(2009)\citenamefont{Chang, Kleban, and
  Levi}}]{Chang_Kleban_Levi:2009}
\bibinfo{author}{\bibfnamefont{S.}~\bibnamefont{Chang}},
  \bibinfo{author}{\bibfnamefont{M.}~\bibnamefont{Kleban}}, \bibnamefont{and}
  \bibinfo{author}{\bibfnamefont{T.~S.} \bibnamefont{Levi}},
  \bibinfo{journal}{JCAP} \textbf{\bibinfo{volume}{0904}}, \bibinfo{pages}{025}
  (\bibinfo{year}{2009}), \eprint{0810.5128}.

\bibitem[{\citenamefont{Chang et~al.}(2008)\citenamefont{Chang, Kleban, and
  Levi}}]{Chang:2007eq}
\bibinfo{author}{\bibfnamefont{S.}~\bibnamefont{Chang}},
  \bibinfo{author}{\bibfnamefont{M.}~\bibnamefont{Kleban}}, \bibnamefont{and}
  \bibinfo{author}{\bibfnamefont{T.~S.} \bibnamefont{Levi}},
  \bibinfo{journal}{JCAP} \textbf{\bibinfo{volume}{0804}}, \bibinfo{pages}{034}
  (\bibinfo{year}{2008}), \eprint{0712.2261}.

\bibitem[{\citenamefont{Czech et~al.}(2010)\citenamefont{Czech, Kleban, Larjo,
  Levi, and Sigurdson}}]{Czech:2010rg}
\bibinfo{author}{\bibfnamefont{B.}~\bibnamefont{Czech}},
  \bibinfo{author}{\bibfnamefont{M.}~\bibnamefont{Kleban}},
  \bibinfo{author}{\bibfnamefont{K.}~\bibnamefont{Larjo}},
  \bibinfo{author}{\bibfnamefont{T.~S.} \bibnamefont{Levi}}, \bibnamefont{and}
  \bibinfo{author}{\bibfnamefont{K.}~\bibnamefont{Sigurdson}},
  \bibinfo{journal}{JCAP} \textbf{\bibinfo{volume}{1012}}, \bibinfo{pages}{023}
  (\bibinfo{year}{2010}), \eprint{1006.0832}.

\bibitem[{\citenamefont{Freivogel et~al.}(2009)\citenamefont{Freivogel, Kleban,
  Nicolis, and Sigurdson}}]{Freivogel_etal:2009it}
\bibinfo{author}{\bibfnamefont{B.}~\bibnamefont{Freivogel}},
  \bibinfo{author}{\bibfnamefont{M.}~\bibnamefont{Kleban}},
  \bibinfo{author}{\bibfnamefont{A.}~\bibnamefont{Nicolis}}, \bibnamefont{and}
  \bibinfo{author}{\bibfnamefont{K.}~\bibnamefont{Sigurdson}},
  \bibinfo{journal}{JCAP} \textbf{\bibinfo{volume}{0908}}, \bibinfo{pages}{036}
  (\bibinfo{year}{2009}).

\bibitem[{\citenamefont{Gobbetti and Kleban}(2012)}]{Gobbetti_Kleban:2012}
\bibinfo{author}{\bibfnamefont{R.}~\bibnamefont{Gobbetti}} \bibnamefont{and}
  \bibinfo{author}{\bibfnamefont{M.}~\bibnamefont{Kleban}},
  \bibinfo{journal}{JCAP} \textbf{\bibinfo{volume}{1205}}, \bibinfo{pages}{025}
  (\bibinfo{year}{2012}), \eprint{1201.6380}.

\bibitem[{\citenamefont{Kleban et~al.}(2011)\citenamefont{Kleban, Levi, and
  Sigurdson}}]{Kleban_Levi_Sigurdson:2011}
\bibinfo{author}{\bibfnamefont{M.}~\bibnamefont{Kleban}},
  \bibinfo{author}{\bibfnamefont{T.~S.} \bibnamefont{Levi}}, \bibnamefont{and}
  \bibinfo{author}{\bibfnamefont{K.}~\bibnamefont{Sigurdson}}
  (\bibinfo{year}{2011}), \eprint{1109.3473}.

\bibitem[{\citenamefont{Kleban}(2011)}]{Kleban:2011pg}
\bibinfo{author}{\bibfnamefont{M.}~\bibnamefont{Kleban}},
  \bibinfo{journal}{Class.Quant.Grav.} \textbf{\bibinfo{volume}{28}},
  \bibinfo{pages}{204008} (\bibinfo{year}{2011}), \eprint{1107.2593}.

\bibitem[{\citenamefont{Wainwright
  et~al.}(2014{\natexlab{a}})\citenamefont{Wainwright, Johnson, Peiris,
  Aguirre, Lehner et~al.}}]{Wainwright:2013lea}
\bibinfo{author}{\bibfnamefont{C.~L.} \bibnamefont{Wainwright}},
  \bibinfo{author}{\bibfnamefont{M.~C.} \bibnamefont{Johnson}},
  \bibinfo{author}{\bibfnamefont{H.~V.} \bibnamefont{Peiris}},
  \bibinfo{author}{\bibfnamefont{A.}~\bibnamefont{Aguirre}},
  \bibinfo{author}{\bibfnamefont{L.}~\bibnamefont{Lehner}},
  \bibnamefont{et~al.}, \bibinfo{journal}{JCAP}
  \textbf{\bibinfo{volume}{1403}}, \bibinfo{pages}{030}
  (\bibinfo{year}{2014}{\natexlab{a}}), \eprint{1312.1357}.

\bibitem[{\citenamefont{Johnson and Yang}(2010)}]{Johnson:2010bn}
\bibinfo{author}{\bibfnamefont{M.~C.} \bibnamefont{Johnson}} \bibnamefont{and}
  \bibinfo{author}{\bibfnamefont{I.-S.} \bibnamefont{Yang}},
  \bibinfo{journal}{Phys.Rev.} \textbf{\bibinfo{volume}{D82}},
  \bibinfo{pages}{065023} (\bibinfo{year}{2010}), \eprint{1005.3506}.

\bibitem[{\citenamefont{Johnson et~al.}(2012)\citenamefont{Johnson, Peiris, and
  Lehner}}]{Johnson:2011wt}
\bibinfo{author}{\bibfnamefont{M.~C.} \bibnamefont{Johnson}},
  \bibinfo{author}{\bibfnamefont{H.~V.} \bibnamefont{Peiris}},
  \bibnamefont{and} \bibinfo{author}{\bibfnamefont{L.}~\bibnamefont{Lehner}},
  \bibinfo{journal}{Phys.Rev.} \textbf{\bibinfo{volume}{D85}},
  \bibinfo{pages}{083516} (\bibinfo{year}{2012}), \eprint{1112.4487}.

\bibitem[{\citenamefont{Freivogel et~al.}(2007)\citenamefont{Freivogel,
  Horowitz, and Shenker}}]{Freivogel:2007fx}
\bibinfo{author}{\bibfnamefont{B.}~\bibnamefont{Freivogel}},
  \bibinfo{author}{\bibfnamefont{G.~T.} \bibnamefont{Horowitz}},
  \bibnamefont{and} \bibinfo{author}{\bibfnamefont{S.}~\bibnamefont{Shenker}}
  (\bibinfo{year}{2007}), \eprint{hep-th/0703146}.

\bibitem[{\citenamefont{Wainwright
  et~al.}(2014{\natexlab{b}})\citenamefont{Wainwright, Johnson, Aguirre, and
  Peiris}}]{Wainwright:2014pta}
\bibinfo{author}{\bibfnamefont{C.~L.} \bibnamefont{Wainwright}},
  \bibinfo{author}{\bibfnamefont{M.~C.} \bibnamefont{Johnson}},
  \bibinfo{author}{\bibfnamefont{A.}~\bibnamefont{Aguirre}}, \bibnamefont{and}
  \bibinfo{author}{\bibfnamefont{H.~V.} \bibnamefont{Peiris}},
  \bibinfo{journal}{JCAP} \textbf{\bibinfo{volume}{1410}}, \bibinfo{pages}{024}
  (\bibinfo{year}{2014}{\natexlab{b}}), \eprint{1407.2950}.

\bibitem[{\citenamefont{Salem et~al.}(2013)\citenamefont{Salem, Saraswat, and
  Shaghoulian}}]{Salem:2012gm}
\bibinfo{author}{\bibfnamefont{M.~P.} \bibnamefont{Salem}},
  \bibinfo{author}{\bibfnamefont{P.}~\bibnamefont{Saraswat}}, \bibnamefont{and}
  \bibinfo{author}{\bibfnamefont{E.}~\bibnamefont{Shaghoulian}},
  \bibinfo{journal}{JCAP} \textbf{\bibinfo{volume}{1302}}, \bibinfo{pages}{019}
  (\bibinfo{year}{2013}), \eprint{1210.4165}.

\bibitem[{\citenamefont{Czech}(2012)}]{Czech:2011aa}
\bibinfo{author}{\bibfnamefont{B.}~\bibnamefont{Czech}},
  \bibinfo{journal}{Phys.Lett.} \textbf{\bibinfo{volume}{B713}},
  \bibinfo{pages}{331} (\bibinfo{year}{2012}), \eprint{1112.1638}.

\bibitem[{\citenamefont{Salem}(2012)}]{Salem:2011qz}
\bibinfo{author}{\bibfnamefont{M.~P.} \bibnamefont{Salem}},
  \bibinfo{journal}{JCAP} \textbf{\bibinfo{volume}{1201}}, \bibinfo{pages}{021}
  (\bibinfo{year}{2012}), \eprint{1108.0040}.

\bibitem[{\citenamefont{Salem}(2010)}]{Salem:2010mi}
\bibinfo{author}{\bibfnamefont{M.~P.} \bibnamefont{Salem}},
  \bibinfo{journal}{Phys.Rev.} \textbf{\bibinfo{volume}{D82}},
  \bibinfo{pages}{063530} (\bibinfo{year}{2010}), \eprint{1005.5311}.

\bibitem[{\citenamefont{Larjo and Levi}(2010)}]{Larjo:2009mt}
\bibinfo{author}{\bibfnamefont{K.}~\bibnamefont{Larjo}} \bibnamefont{and}
  \bibinfo{author}{\bibfnamefont{T.~S.} \bibnamefont{Levi}},
  \bibinfo{journal}{JCAP} \textbf{\bibinfo{volume}{1008}}, \bibinfo{pages}{034}
  (\bibinfo{year}{2010}), \eprint{0910.4159}.

\bibitem[{\citenamefont{Easther et~al.}(2009)\citenamefont{Easther, Giblin,
  Hui, and Lim}}]{Easther:2009ft}
\bibinfo{author}{\bibfnamefont{R.}~\bibnamefont{Easther}},
  \bibinfo{author}{\bibfnamefont{J.}~\bibnamefont{Giblin},
  \bibfnamefont{John~T.}},
  \bibinfo{author}{\bibfnamefont{L.}~\bibnamefont{Hui}}, \bibnamefont{and}
  \bibinfo{author}{\bibfnamefont{E.~A.} \bibnamefont{Lim}},
  \bibinfo{journal}{Phys.Rev.} \textbf{\bibinfo{volume}{D80}},
  \bibinfo{pages}{123519} (\bibinfo{year}{2009}), \eprint{0907.3234}.

\bibitem[{\citenamefont{Giblin et~al.}(2010)\citenamefont{Giblin, Hui, Lim, and
  Yang}}]{Giblin:2010bd}
\bibinfo{author}{\bibfnamefont{J.}~\bibnamefont{Giblin},
  \bibfnamefont{John~T.}},
  \bibinfo{author}{\bibfnamefont{L.}~\bibnamefont{Hui}},
  \bibinfo{author}{\bibfnamefont{E.~A.} \bibnamefont{Lim}}, \bibnamefont{and}
  \bibinfo{author}{\bibfnamefont{I.-S.} \bibnamefont{Yang}},
  \bibinfo{journal}{Phys.Rev.} \textbf{\bibinfo{volume}{D82}},
  \bibinfo{pages}{045019} (\bibinfo{year}{2010}), \eprint{1005.3493}.

\bibitem[{\citenamefont{Ahlqvist et~al.}(2014)\citenamefont{Ahlqvist, Eckerle,
  and Greene}}]{Ahlqvist:2014uha}
\bibinfo{author}{\bibfnamefont{P.}~\bibnamefont{Ahlqvist}},
  \bibinfo{author}{\bibfnamefont{K.}~\bibnamefont{Eckerle}}, \bibnamefont{and}
  \bibinfo{author}{\bibfnamefont{B.}~\bibnamefont{Greene}}
  (\bibinfo{year}{2014}), \eprint{1411.4631}.

\bibitem[{\citenamefont{Kim et~al.}(2014)\citenamefont{Kim, Lee, Lee, Yang, and
  Yeom}}]{Kim:2014ara}
\bibinfo{author}{\bibfnamefont{D.-H.} \bibnamefont{Kim}},
  \bibinfo{author}{\bibfnamefont{B.-H.} \bibnamefont{Lee}},
  \bibinfo{author}{\bibfnamefont{W.}~\bibnamefont{Lee}},
  \bibinfo{author}{\bibfnamefont{J.}~\bibnamefont{Yang}}, \bibnamefont{and}
  \bibinfo{author}{\bibfnamefont{D.-h.} \bibnamefont{Yeom}}
  (\bibinfo{year}{2014}), \eprint{1410.4648}.

\bibitem[{\citenamefont{Ahlqvist et~al.}(2013)\citenamefont{Ahlqvist, Eckerle,
  and Greene}}]{Ahlqvist:2013whn}
\bibinfo{author}{\bibfnamefont{P.}~\bibnamefont{Ahlqvist}},
  \bibinfo{author}{\bibfnamefont{K.}~\bibnamefont{Eckerle}}, \bibnamefont{and}
  \bibinfo{author}{\bibfnamefont{B.}~\bibnamefont{Greene}}
  (\bibinfo{year}{2013}), \eprint{1310.6069}.

\bibitem[{\citenamefont{Hwang et~al.}(2012)\citenamefont{Hwang, Lee, Lee, and
  Yeom}}]{Hwang:2012pj}
\bibinfo{author}{\bibfnamefont{D.-i.} \bibnamefont{Hwang}},
  \bibinfo{author}{\bibfnamefont{B.-H.} \bibnamefont{Lee}},
  \bibinfo{author}{\bibfnamefont{W.}~\bibnamefont{Lee}}, \bibnamefont{and}
  \bibinfo{author}{\bibfnamefont{D.-h.} \bibnamefont{Yeom}},
  \bibinfo{journal}{JCAP} \textbf{\bibinfo{volume}{1207}}, \bibinfo{pages}{003}
  (\bibinfo{year}{2012}), \eprint{1201.6109}.

\bibitem[{\citenamefont{Deskins et~al.}(2012)\citenamefont{Deskins, Giblin, and
  Yang}}]{Deskins:2012tj}
\bibinfo{author}{\bibfnamefont{J.~T.} \bibnamefont{Deskins}},
  \bibinfo{author}{\bibfnamefont{J.}~\bibnamefont{Giblin},
  \bibfnamefont{John~T.}}, \bibnamefont{and}
  \bibinfo{author}{\bibfnamefont{I.-S.} \bibnamefont{Yang}},
  \bibinfo{journal}{JHEP} \textbf{\bibinfo{volume}{1210}}, \bibinfo{pages}{035}
  (\bibinfo{year}{2012}), \eprint{1207.6636}.

\bibitem[{\citenamefont{Amin et~al.}(2013{\natexlab{a}})\citenamefont{Amin,
  Lim, and Yang}}]{Amin:2013dqa}
\bibinfo{author}{\bibfnamefont{M.~A.} \bibnamefont{Amin}},
  \bibinfo{author}{\bibfnamefont{E.~A.} \bibnamefont{Lim}}, \bibnamefont{and}
  \bibinfo{author}{\bibfnamefont{I.-S.} \bibnamefont{Yang}}
  (\bibinfo{year}{2013}{\natexlab{a}}), \eprint{1308.0605}.

\bibitem[{\citenamefont{Amin et~al.}(2013{\natexlab{b}})\citenamefont{Amin,
  Lim, and Yang}}]{Amin:2013eqa}
\bibinfo{author}{\bibfnamefont{M.~A.} \bibnamefont{Amin}},
  \bibinfo{author}{\bibfnamefont{E.~A.} \bibnamefont{Lim}}, \bibnamefont{and}
  \bibinfo{author}{\bibfnamefont{I.-S.} \bibnamefont{Yang}}
  (\bibinfo{year}{2013}{\natexlab{b}}), \eprint{1308.0606}.

\bibitem[{\citenamefont{Feeney et~al.}(2011{\natexlab{a}})\citenamefont{Feeney,
  Johnson, Mortlock, and Peiris}}]{Feeney_etal:2010dd}
\bibinfo{author}{\bibfnamefont{S.~M.} \bibnamefont{Feeney}},
  \bibinfo{author}{\bibfnamefont{M.~C.} \bibnamefont{Johnson}},
  \bibinfo{author}{\bibfnamefont{D.~J.} \bibnamefont{Mortlock}},
  \bibnamefont{and} \bibinfo{author}{\bibfnamefont{H.~V.}
  \bibnamefont{Peiris}}, \bibinfo{journal}{Phys. Rev.}
  \textbf{\bibinfo{volume}{D84}}, \bibinfo{pages}{043507}
  (\bibinfo{year}{2011}{\natexlab{a}}), \eprint{1012.3667}.

\bibitem[{\citenamefont{Feeney et~al.}(2011{\natexlab{b}})\citenamefont{Feeney,
  Johnson, Mortlock, and Peiris}}]{Feeney_etal:2010jj}
\bibinfo{author}{\bibfnamefont{S.~M.} \bibnamefont{Feeney}},
  \bibinfo{author}{\bibfnamefont{M.~C.} \bibnamefont{Johnson}},
  \bibinfo{author}{\bibfnamefont{D.~J.} \bibnamefont{Mortlock}},
  \bibnamefont{and} \bibinfo{author}{\bibfnamefont{H.~V.}
  \bibnamefont{Peiris}}, \bibinfo{journal}{Phys. Rev. Lett.}
  \textbf{\bibinfo{volume}{107}}, \bibinfo{pages}{071301}
  (\bibinfo{year}{2011}{\natexlab{b}}), \eprint{1012.1995}.

\bibitem[{\citenamefont{Feeney et~al.}(2013)\citenamefont{Feeney, Johnson,
  McEwen, Mortlock, and Peiris}}]{Feeney:2012hj}
\bibinfo{author}{\bibfnamefont{S.~M.} \bibnamefont{Feeney}},
  \bibinfo{author}{\bibfnamefont{M.~C.} \bibnamefont{Johnson}},
  \bibinfo{author}{\bibfnamefont{J.~D.} \bibnamefont{McEwen}},
  \bibinfo{author}{\bibfnamefont{D.~J.} \bibnamefont{Mortlock}},
  \bibnamefont{and} \bibinfo{author}{\bibfnamefont{H.~V.}
  \bibnamefont{Peiris}}, \bibinfo{journal}{Phys.Rev.}
  \textbf{\bibinfo{volume}{D88}}, \bibinfo{pages}{043012}
  (\bibinfo{year}{2013}), \eprint{1210.2725}.

\bibitem[{\citenamefont{McEwen et~al.}(2012)\citenamefont{McEwen, Feeney,
  Johnson, and Peiris}}]{McEwen:2012uk}
\bibinfo{author}{\bibfnamefont{J.~D.} \bibnamefont{McEwen}},
  \bibinfo{author}{\bibfnamefont{S.~M.} \bibnamefont{Feeney}},
  \bibinfo{author}{\bibfnamefont{M.~C.} \bibnamefont{Johnson}},
  \bibnamefont{and} \bibinfo{author}{\bibfnamefont{H.~V.}
  \bibnamefont{Peiris}}, \bibinfo{journal}{Phys. Rev.}
  \textbf{\bibinfo{volume}{D85}}, \bibinfo{pages}{103502}
  (\bibinfo{year}{2012}), \eprint{1202.2861}.

\bibitem[{\citenamefont{Osborne
  et~al.}(2013{\natexlab{a}})\citenamefont{Osborne, Senatore, and
  Smith}}]{Osborne:2013hea}
\bibinfo{author}{\bibfnamefont{S.}~\bibnamefont{Osborne}},
  \bibinfo{author}{\bibfnamefont{L.}~\bibnamefont{Senatore}}, \bibnamefont{and}
  \bibinfo{author}{\bibfnamefont{K.}~\bibnamefont{Smith}}
  (\bibinfo{year}{2013}{\natexlab{a}}), \eprint{1305.1964}.

\bibitem[{\citenamefont{Osborne
  et~al.}(2013{\natexlab{b}})\citenamefont{Osborne, Senatore, and
  Smith}}]{Osborne:2013jea}
\bibinfo{author}{\bibfnamefont{S.}~\bibnamefont{Osborne}},
  \bibinfo{author}{\bibfnamefont{L.}~\bibnamefont{Senatore}}, \bibnamefont{and}
  \bibinfo{author}{\bibfnamefont{K.}~\bibnamefont{Smith}}
  (\bibinfo{year}{2013}{\natexlab{b}}), \eprint{1305.1970}.

\bibitem[{\citenamefont{{Planck Collaboration}
  et~al.}(2014)\citenamefont{{Planck Collaboration}, {Ade}, {Aghanim},
  {Armitage-Caplan}, {Arnaud}, {Ashdown}, {Atrio-Barandela}, {Aumont},
  {Baccigalupi}, {Banday} et~al.}}]{2014A&A...571A..16P}
\bibinfo{author}{\bibnamefont{{Planck Collaboration}}},
  \bibinfo{author}{\bibfnamefont{P.~A.~R.} \bibnamefont{{Ade}}},
  \bibinfo{author}{\bibfnamefont{N.}~\bibnamefont{{Aghanim}}},
  \bibinfo{author}{\bibfnamefont{C.}~\bibnamefont{{Armitage-Caplan}}},
  \bibinfo{author}{\bibfnamefont{M.}~\bibnamefont{{Arnaud}}},
  \bibinfo{author}{\bibfnamefont{M.}~\bibnamefont{{Ashdown}}},
  \bibinfo{author}{\bibfnamefont{F.}~\bibnamefont{{Atrio-Barandela}}},
  \bibinfo{author}{\bibfnamefont{J.}~\bibnamefont{{Aumont}}},
  \bibinfo{author}{\bibfnamefont{C.}~\bibnamefont{{Baccigalupi}}},
  \bibinfo{author}{\bibfnamefont{A.~J.} \bibnamefont{{Banday}}},
  \bibnamefont{et~al.}, \bibinfo{journal}{Astron. Astrophys.}
  \textbf{\bibinfo{volume}{571}}, \bibinfo{eid}{A16} (\bibinfo{year}{2014}),
  \eprint{1303.5076},
  \urlprefix\url{http://adsabs.harvard.edu/abs/2014A%26A...571A..16P}.

\bibitem[{\citenamefont{{Komatsu} et~al.}(2003)\citenamefont{{Komatsu},
  {Kogut}, {Nolta}, {Bennett}, {Halpern}, {Hinshaw}, {Jarosik}, {Limon},
  {Meyer}, {Page} et~al.}}]{2003ApJS..148..119K}
\bibinfo{author}{\bibfnamefont{E.}~\bibnamefont{{Komatsu}}},
  \bibinfo{author}{\bibfnamefont{A.}~\bibnamefont{{Kogut}}},
  \bibinfo{author}{\bibfnamefont{M.~R.} \bibnamefont{{Nolta}}},
  \bibinfo{author}{\bibfnamefont{C.~L.} \bibnamefont{{Bennett}}},
  \bibinfo{author}{\bibfnamefont{M.}~\bibnamefont{{Halpern}}},
  \bibinfo{author}{\bibfnamefont{G.}~\bibnamefont{{Hinshaw}}},
  \bibinfo{author}{\bibfnamefont{N.}~\bibnamefont{{Jarosik}}},
  \bibinfo{author}{\bibfnamefont{M.}~\bibnamefont{{Limon}}},
  \bibinfo{author}{\bibfnamefont{S.~S.} \bibnamefont{{Meyer}}},
  \bibinfo{author}{\bibfnamefont{L.}~\bibnamefont{{Page}}},
  \bibnamefont{et~al.}, \bibinfo{journal}{Astrophys. J. Suppl. Ser.}
  \textbf{\bibinfo{volume}{148}}, \bibinfo{pages}{119} (\bibinfo{year}{2003}),
  \eprint{astro-ph/0302223},
  \urlprefix\url{http://adsabs.harvard.edu/cgi-bin/nph-bib_query?bibcode=2003ApJS..148..119K&db_key=AST}.

\bibitem[{\citenamefont{{Lewis} and {Bridle}}(2002)}]{2002PhRvD..66j3511L}
\bibinfo{author}{\bibfnamefont{A.}~\bibnamefont{{Lewis}}} \bibnamefont{and}
  \bibinfo{author}{\bibfnamefont{S.}~\bibnamefont{{Bridle}}},
  \bibinfo{journal}{Phys. Rev. D} \textbf{\bibinfo{volume}{66}},
  \bibinfo{pages}{103511} (\bibinfo{year}{2002}),
  \eprint{arXiv:astro-ph/0205436},
  \urlprefix\url{http://adsabs.harvard.edu/abs/2002PhRvD..66j3511L}.

\bibitem[{\citenamefont{{Elsner} and {Wandelt}}(2009)}]{2009ApJS..184..264E}
\bibinfo{author}{\bibfnamefont{F.}~\bibnamefont{{Elsner}}} \bibnamefont{and}
  \bibinfo{author}{\bibfnamefont{B.~D.} \bibnamefont{{Wandelt}}},
  \bibinfo{journal}{Astrophys. J. Suppl. Ser.} \textbf{\bibinfo{volume}{184}},
  \bibinfo{pages}{264} (\bibinfo{year}{2009}), \eprint{0909.0009},
  \urlprefix\url{http://adsabs.harvard.edu/abs/2009ApJS..184..264E}.

\bibitem[{\citenamefont{{Galli} et~al.}(2014)\citenamefont{{Galli}, {Benabed},
  {Bouchet}, {Cardoso}, {Elsner}, {Hivon}, {Mangilli}, {Prunet}, and
  {Wandelt}}}]{2014PhRvD..90f3504G}
\bibinfo{author}{\bibfnamefont{S.}~\bibnamefont{{Galli}}},
  \bibinfo{author}{\bibfnamefont{K.}~\bibnamefont{{Benabed}}},
  \bibinfo{author}{\bibfnamefont{F.}~\bibnamefont{{Bouchet}}},
  \bibinfo{author}{\bibfnamefont{J.-F.} \bibnamefont{{Cardoso}}},
  \bibinfo{author}{\bibfnamefont{F.}~\bibnamefont{{Elsner}}},
  \bibinfo{author}{\bibfnamefont{E.}~\bibnamefont{{Hivon}}},
  \bibinfo{author}{\bibfnamefont{A.}~\bibnamefont{{Mangilli}}},
  \bibinfo{author}{\bibfnamefont{S.}~\bibnamefont{{Prunet}}}, \bibnamefont{and}
  \bibinfo{author}{\bibfnamefont{B.}~\bibnamefont{{Wandelt}}},
  \bibinfo{journal}{Phys.Rev.} \textbf{\bibinfo{volume}{D90}},
  \bibinfo{eid}{063504} (\bibinfo{year}{2014}), \eprint{1403.5271}.

\bibitem[{\citenamefont{{Jarosik} et~al.}(2011)\citenamefont{{Jarosik},
  {Bennett}, {Dunkley}, {Gold}, {Greason}, {Halpern}, {Hill}, {Hinshaw},
  {Kogut}, {Komatsu} et~al.}}]{WMAP7_Jarosik}
\bibinfo{author}{\bibfnamefont{N.}~\bibnamefont{{Jarosik}}},
  \bibinfo{author}{\bibfnamefont{C.~L.} \bibnamefont{{Bennett}}},
  \bibinfo{author}{\bibfnamefont{J.}~\bibnamefont{{Dunkley}}},
  \bibinfo{author}{\bibfnamefont{B.}~\bibnamefont{{Gold}}},
  \bibinfo{author}{\bibfnamefont{M.~R.} \bibnamefont{{Greason}}},
  \bibinfo{author}{\bibfnamefont{M.}~\bibnamefont{{Halpern}}},
  \bibinfo{author}{\bibfnamefont{R.~S.} \bibnamefont{{Hill}}},
  \bibinfo{author}{\bibfnamefont{G.}~\bibnamefont{{Hinshaw}}},
  \bibinfo{author}{\bibfnamefont{A.}~\bibnamefont{{Kogut}}},
  \bibinfo{author}{\bibfnamefont{E.}~\bibnamefont{{Komatsu}}},
  \bibnamefont{et~al.}, \bibinfo{journal}{Astrophys. J. Suppl. Ser.}
  \textbf{\bibinfo{volume}{192}}, \bibinfo{eid}{14} (\bibinfo{year}{2011}),
  \eprint{1001.4744}.

\bibitem[{\citenamefont{{Tegmark} et~al.}(1997)\citenamefont{{Tegmark},
  {Taylor}, and {Heavens}}}]{1997ApJ...480...22T}
\bibinfo{author}{\bibfnamefont{M.}~\bibnamefont{{Tegmark}}},
  \bibinfo{author}{\bibfnamefont{A.~N.} \bibnamefont{{Taylor}}},
  \bibnamefont{and} \bibinfo{author}{\bibfnamefont{A.~F.}
  \bibnamefont{{Heavens}}}, \bibinfo{journal}{Astrophys. J.}
  \textbf{\bibinfo{volume}{480}}, \bibinfo{pages}{22} (\bibinfo{year}{1997}),
  \eprint{astro-ph/9603021}.

\bibitem[{\citenamefont{Zhang and Johnson}(2015)}]{Zhang:2015uta}
\bibinfo{author}{\bibfnamefont{P.}~\bibnamefont{Zhang}} \bibnamefont{and}
  \bibinfo{author}{\bibfnamefont{M.~C.} \bibnamefont{Johnson}}
  (\bibinfo{year}{2015}), \eprint{1501.00511}.

\bibitem[{\citenamefont{{LSST Science Collaboration}
  et~al.}(2009)\citenamefont{{LSST Science Collaboration}, {Abell}, {Allison},
  {Anderson}, {Andrew}, {Angel}, {Armus}, {Arnett}, {Asztalos}, {Axelrod}
  et~al.}}]{2009arXiv0912.0201L}
\bibinfo{author}{\bibnamefont{{LSST Science Collaboration}}},
  \bibinfo{author}{\bibfnamefont{P.~A.} \bibnamefont{{Abell}}},
  \bibinfo{author}{\bibfnamefont{J.}~\bibnamefont{{Allison}}},
  \bibinfo{author}{\bibfnamefont{S.~F.} \bibnamefont{{Anderson}}},
  \bibinfo{author}{\bibfnamefont{J.~R.} \bibnamefont{{Andrew}}},
  \bibinfo{author}{\bibfnamefont{J.~R.~P.} \bibnamefont{{Angel}}},
  \bibinfo{author}{\bibfnamefont{L.}~\bibnamefont{{Armus}}},
  \bibinfo{author}{\bibfnamefont{D.}~\bibnamefont{{Arnett}}},
  \bibinfo{author}{\bibfnamefont{S.~J.} \bibnamefont{{Asztalos}}},
  \bibinfo{author}{\bibfnamefont{T.~S.} \bibnamefont{{Axelrod}}},
  \bibnamefont{et~al.}, \bibinfo{journal}{ArXiv e-prints}
  (\bibinfo{year}{2009}), \eprint{0912.0201}.

\bibitem[{\citenamefont{Braden et~al.}(2015{\natexlab{a}})\citenamefont{Braden,
  Bond, and Mersini-Houghton}}]{Braden:2014cra}
\bibinfo{author}{\bibfnamefont{J.}~\bibnamefont{Braden}},
  \bibinfo{author}{\bibfnamefont{J.~R.} \bibnamefont{Bond}}, \bibnamefont{and}
  \bibinfo{author}{\bibfnamefont{L.}~\bibnamefont{Mersini-Houghton}},
  \bibinfo{journal}{JCAP} \textbf{\bibinfo{volume}{1503}}, \bibinfo{pages}{007}
  (\bibinfo{year}{2015}{\natexlab{a}}), \eprint{1412.5591}.

\bibitem[{\citenamefont{Braden et~al.}(2015{\natexlab{b}})\citenamefont{Braden,
  Bond, and Mersini-Houghton}}]{Braden:2015vza}
\bibinfo{author}{\bibfnamefont{J.}~\bibnamefont{Braden}},
  \bibinfo{author}{\bibfnamefont{J.~R.} \bibnamefont{Bond}}, \bibnamefont{and}
  \bibinfo{author}{\bibfnamefont{L.}~\bibnamefont{Mersini-Houghton}}
  (\bibinfo{year}{2015}{\natexlab{b}}), \eprint{1505.01857}.

\bibitem[{\citenamefont{Bond et~al.}(2015)\citenamefont{Bond, Braden, and
  Mersini-Houghton}}]{Bond:2015zfa}
\bibinfo{author}{\bibfnamefont{J.~R.} \bibnamefont{Bond}},
  \bibinfo{author}{\bibfnamefont{J.}~\bibnamefont{Braden}}, \bibnamefont{and}
  \bibinfo{author}{\bibfnamefont{L.}~\bibnamefont{Mersini-Houghton}}
  (\bibinfo{year}{2015}), \eprint{1505.02162}.

\end{thebibliography}

\end{document}